\documentclass[prd,nofootinbib,preprint,superscriptaddress]{revtex4}
\pdfoutput=1
\usepackage[T1]{fontenc}
\usepackage{amsmath,amssymb}
\usepackage{epsfig}
\usepackage{graphicx}
\usepackage[usenames,dvipsnames]{color}
\usepackage{subfigure}
\usepackage{slashed}
\usepackage[colorlinks,citecolor=blue]{hyperref}
\usepackage{pdfpages}
\usepackage{color}
\usepackage{comment}

\begin{document}
\title{Leptogenesis and Dark Matter Through Relativistic Bubble Walls with Observable Gravitational Waves}

\author{Debasish Borah}
\email{dborah@iitg.ac.in}
\affiliation{Department of Physics, Indian Institute of Technology Guwahati, Assam 781039, India}

\author{Arnab Dasgupta}
\email{arnabdasgupta@pitt.edu}
\affiliation{Pittsburgh Particle Physics, Astrophysics, and Cosmology Center, Department of Physics and Astronomy, University of Pittsburgh, Pittsburgh, PA 15206, USA}

\author{Indrajit Saha}
\email{s.indrajit@iitg.ac.in}
\affiliation{Department of Physics, Indian Institute of Technology Guwahati, Assam 781039, India}

\begin{abstract}
We study a scenario where both dark matter and heavy right handed neutrino (RHN) responsible for leptogenesis acquire masses by crossing the relativistic bubble walls formed as a result of a TeV scale supercooled first order phase transition (FOPT). While this leads to a large out-of-equilibrium abundance of right handed neutrino inside the bubble sufficient to produce the required lepton asymmetry, the dark matter being lighter can still remain in equilibrium with its relic being set by subsequent thermal freeze-out. A classical conformal symmetry ensures the origin of mass via FOPT induced by a singlet scalar while also ensuring supercooling leading to enhanced gravitational wave amplitude within the sensitivity of the LISA experiment. A minimal scenario with three RHN, one inert scalar doublet and one singlet scalar as additional fields beyond the standard model is sufficient to realize this possibility which also favours inert RHN dark matter over inert scalar doublet.
\end{abstract}

\maketitle

\section{Introduction}
The observed baryon asymmetry of the universe (BAU) has been a longstanding puzzle in cosmology and particle physics. While only around $5\%$ of the present universe is made up of baryonic matter, the observed excess of baryons over anti-baryons is quoted in terms of the baryon to photon ratio as \cite{Aghanim:2018eyx} 
\begin{equation}
\eta_B = \frac{n_{B}-n_{\overline{B}}}{n_{\gamma}} \simeq 6.2 \times 10^{-10}, 
\label{etaBobs}
\end{equation} 
based on the cosmic microwave background (CMB) measurements which also agrees well with the big bang nucleosynthesis (BBN) estimates \cite{Zyla:2020zbs}. Assuming the universe to start in a matter-antimatter symmetric manner, the observed BAU can be generated dynamically if certain conditions, known as the Sakharov's conditions \cite{Sakharov:1967dj} are satisfied. Since the standard model (SM) fails to satisfy these criteria in required amount, several ways of generating the BAU has been proposed. In one such popular scenario, known as baryogenesis \cite{Weinberg:1979bt, Kolb:1979qa}, out-of-equilibrium decay of a heavy particle is responsible for generating the observed BAU. One interesting way to achieve baryogenesis is leptogenesis \cite{Fukugita:1986hr} where a non-zero lepton asymmetry is first generated which later gets converted into the BAU via electroweak sphalerons \cite{Kuzmin:1985mm}. While this asymmetric baryonic matter comprises $5\%$ of the present universe, the total matter content of the universe is around $32\%$ with the significant portion being in the form of a non-luminous, non-baryonic form of matter, known as dark matter (DM). While relative abundance of DM is approximately $27\%$, it is conventionally reported in terms of density parameter $\Omega_{\rm DM}$ and reduced Hubble constant $h = \text{Hubble Parameter}/(100 \;\text{km} ~\text{s}^{-1} 
\text{Mpc}^{-1})$ as \cite{Aghanim:2018eyx}
\begin{equation}
\Omega_{\text{DM}} h^2 = 0.120\pm 0.001
\label{dm_relic}
\end{equation}
\noindent at 68\% CL. Similar to BAU, there is no explanation for DM in the SM, leading to a plethora of beyond standard model (BSM) scenarios. The weakly interacting massive particle (WIMP) paradigm has been the most widely studied particle DM scenario where a DM particle having mass and interactions around the electroweak ballpark naturally gives rise to the observed DM relic via thermal freeze-out, a remarkable coincidence often referred to as the {\it WIMP Miracle}.

In generic seesaw scenarios, there exists a lower bound on the scale of leptogenesis $M_1 > 10^9$ GeV, known as the Davidson-Ibarra bound \cite{Davidson:2002qv} if such asymmetries arise from out-of-equilibrium decay\footnote{However, it is worth mentioning that, lepton asymmetry can also be generated from oscillations \cite{Akhmedov:1998qx, Asaka:2005pn, Abada:2018oly, Drewes:2021nqr} where the scale of leptogenesis in minimal seesaw model can be as low as sub-GeV scale.}. This keeps the scale of leptogenesis far away from any direct experimental probe. In scenarios where lepton asymmetry is generated from decay, introduction of additional fields on top of the ones required to implement a generic seesaw model of neutrino mass, can alleviate such strong lower bound on the scale of leptogenesis \cite{LeDall:2014too, Alanne:2018brf, Hambye:2009pw, Racker:2013lua, Clarke:2015hta, Hugle:2018qbw, Borah:2018rca, Mahanta:2019gfe, Mahanta:2019sfo, Sarma:2020msa, Borah:2020ivi}. Even in such leptogenesis from decay type scenarios, there is another way to have TeV scale leptogenesis by resonant enhancement of the CP asymmetry, known as the resonant leptogenesis \cite{Pilaftsis:1998pd, Pilaftsis:2003gt} with fine-tuned mass splitting between decaying particles. Even for such TeV scale leptogenesis, the decaying particle say, a right handed neutrino, has very tiny couplings with leptons in order to satisfy light neutrino masses, making it difficult to probe it directly. Thus, leptogenesis from decay, in general, has limited direct experimental probe \cite{Chun:2017spz}. This has led to some recent attempts in finding ways to probe leptogenesis via stochastic gravitational wave (GW) observation \cite{Dror:2019syi, Blasi:2020wpy, Fornal:2020esl, Samanta:2020cdk, Barman:2022yos, Baldes:2021vyz, Azatov:2021irb, Huang:2022vkf, Dasgupta:2022isg}. While some of these works rely upon topological defects like cosmic strings, domain walls formed as a result of symmetry breaking \cite{Dror:2019syi, Blasi:2020wpy, Fornal:2020esl, Samanta:2020cdk, Barman:2022yos}, others consider a first order phase transition (FOPT) to be responsible for generating GW \cite{Baldes:2021vyz, Azatov:2021irb, Huang:2022vkf, Dasgupta:2022isg}. Similarly, there have been attempts in finding complementary DM probes like stochastic GW background \cite{Yuan:2021ebu, Tsukada:2020lgt, Chatrchyan:2020pzh, Bian:2021vmi, Samanta:2021mdm, Borah:2022byb, Azatov:2021ifm, Azatov:2022tii, Baldes:2022oev}, specially in view of the continued null results at direct detection experiments \cite{LUX-ZEPLIN:2022qhg}.

Motivated by this, we consider a minimal scenario where both leptogenesis and DM are triggered by a strong first order phase transition with observable GW signatures. Similar to the baryogenesis mechanism adopted in \cite{Baldes:2021vyz} followed by leptogenesis implementation in \cite{Huang:2022vkf, Dasgupta:2022isg}, we consider a scenario where DM as well as right handed neutrino responsible for leptogenesis acquire masses in a FOPT by crossing the relativistic bubble walls\footnote{See \cite{Arakawa:2021wgz, Ahmadvand:2021vxs} for other scenarios connecting FOPT to baryogenesis.}. Unlike additional gauge symmetries considered in these works, here we consider a minimal scenario without any gauge extension of the SM. Adopting a classical conformal symmetry required to generate masses of gauge singlet fermions via FOPT as well as to enhance the strength via supercooling, we first consider the conformal version of the minimal scotogenic model \cite{Ma:2006km}. While this model contains DM as well as right handed neutrino (RHN) responsible for leptogenesis, we find that it is not possible to get the desired leptogenesis from RHN decay. This is due to strict constraints on the hierarchy of massive fields like DM and RHNs coupling to the singlet scalar field which is also driving the FOPT. We then adopt a hybrid setup with type I and scotogenic origin of light neutrino masses and discuss the resulting phenomenology of DM, leptogenesis and GW. With only five additional BSM fields, the model remains successful and predictive at experiments operational at different frontiers.

This paper is organised as follows. In section \ref{sec2} we briefly discuss the conformal version of the scotogenic model followed by the details of first order phase transition in section \ref{sec3}. In section \ref{sec4}, we discuss the details of gravitational wave production followed by discussion of leptogenesis and dark matter in section \ref{sec5}. Finally, we conclude in section \ref{sec6}.

\section{Conformal Scotogenic Model}
\label{sec2}
In order to realize a supercooled phase transition along with leptogenesis and dark matter, we first consider the conformal or scale invariant version \cite{Ahriche:2016cio} of the minimal scotogenic model \cite{Ma:2006km} where the SM is extended by three gauge singlet right handed neutrinos $N_i$ (with $i=1,2,3$), one additional scalar doublet $\eta$. An additional $Z_2$ symmetry is imposed under which these newly added particles are odd while all SM particles are even. In order to preserve the conformal nature and generate masses, an additional $Z_2$ even singlet scalar S is introduced.

The leptonic Yukawa Lagrangian relevant for light neutrino mass is
\begin{equation}\label{IRHYukawa}
{\cal L} \ \supset \ \frac{1}{2}Y'_{ij} S N_iN_j + \left(Y_{ij} \, \bar{L}_i \tilde{\eta} N_j  + \text{h.c.} \right) \ . 
\end{equation}
Clearly, there is no coupling of neutrinos to the SM Higgs doublet $\Phi_1$ due to the unbroken $Z_2$ symmetry. However, light neutrino masses arise at radiative level with $Z_2$ odd particles taking part in the loop. 

The scalar potential of the model can be written as
\begin{align}
V(\Phi_1,\eta, S) & \ = \ \frac{\lambda_1}{4}|\Phi_1|^4+\frac{\lambda_2}{4}|\eta|^4+\lambda_3|\Phi_1|^2|\eta|^2 + \frac{1}{4} \lambda_S S^4 +\lambda_4|\Phi_1^\dag \eta|^2 + \left[\frac{\lambda_5}{2}(\Phi_1^\dag \eta)^2 + \text{h.c.}\right] \nonumber \\
& \qquad +\lambda_6|\Phi_1|^2 S^2 + \lambda_7 |\eta|^2 S^2 \, . \label{eq:tree potential}
\end{align}
where $\Phi_1$ is the SM Higgs doublet. Light neutrino masses which arise at one loop level can be evaluated as ~\cite{Ma:2006km, Merle:2015ica}
\begin{align}
(m_{\nu})_{ij} \ & = \ \sum_k \frac{Y_{ik}Y_{jk} M_{k}}{32 \pi^2} \left ( \frac{m^2_{H^0}}{m^2_{H^0}-M^2_k} \: \text{ln} \frac{m^2_{H^0}}{M^2_k}-\frac{m^2_{A^0}}{m^2_{A^0}-M^2_k}\: \text{ln} \frac{m^2_{A^0}}{M^2_k} \right) \nonumber \\ 
& \ \equiv  \ \sum_k \frac{Y_{ik}Y_{jk} M_{k}}{32 \pi^2} \left[L_k(m^2_{H^0})-L_k(m^2_{A^0})\right] \, ,
\label{numass1}
\end{align}
where 
$M_k$ is the mass eigenvalue of the mass eigenstate $N_k$ in the internal line and the indices $i, j = 1,2,3$ run over the three neutrino generations as well as three copies of $N_i$. Also, $A^0, H^0$ are the neutral pseudoscalar and scalar respectively contained in $\eta$. The function $L_k(m^2)$ is defined as 
\begin{align}
L_k(m^2) \ = \ \frac{m^2}{m^2-M^2_k} \: \text{ln} \frac{m^2}{M^2_k} \, .
\label{eq:Lk}
\end{align}
Using the physical scalar mass expressions \cite{Mahanta:2019sfo}, one can write $m^2_{H^0}-m^2_{A^0}=\lambda_5 v^2_{\rm ew}$ where $v_{\rm ew}$ is the vacuum expectation value (VEV) of SM Higgs doublet. Thus, light neutrino mass is directly proportional to the parameter $\lambda_5$. In upcoming discussions, we will discuss the effects of $\lambda_5$ in details.

\section{First order phase transition}
\label{sec3}
Here we are interested in the phase transition driven by the singlet scalar field at a scale above the electroweak scale as preferred from leptogenesis point of view. For electroweak phase transition in minimal scotogenic model without any conformal symmetry or singlet scalar, please refer to earlier work \cite{Borah:2020wut}.

In order to study the details of FOPT in conformal scotogenic model, we first write down the full one-loop potential which can be schematically divided into following form:
\begin{align}
V_{\rm tot} = V_{\rm tree} + V_{\rm CW} +V_{\rm th},
\end{align}
where $V_{\rm tree},~V_{\rm CW}$ and $V_{\rm th}$ denote the tree level scalar potential, the one-loop Coleman-Weinberg potential, the thermal effective potential, respectively.
The tree level scalar potential is given by Eq.~\eqref{eq:tree potential}.
In finite-temperature field theory, the effective potential, $V_{\rm CW}$ and $V_{\rm thermal}$, are calculated by using standard background field method~\cite{Dolan:1973qd,Quiros:1999jp}.
In the following calculations, we take Landau gauge for simplicity.\footnote{The gauge dependence of the thermal effective potential is discussed by many authors. See. e.g. Refs.~\cite{Wainwright:2011qy,Wainwright:2012zn} and references therein.}

The Coleman-Weinberg potential~\cite{Coleman:1973jx} with $\overline{\rm DR}$ regularisation is given by
\begin{align}
V_{\rm CW} = \sum_i (-)^{n_{f}} \frac{n_i}{64\pi^2} m_i^4 (\phi) \left(\log\left(\frac{m_i^2 (\phi)}{\mu^2} \right)-\frac{3}{2} \right),
\end{align}
where suffix $i$ represents particle species, and $n_i,~m_i (\phi)$ are the degrees of freedom (dof) and field dependent masses of $i$'th particle.
In addition, $\mu$ is the renormalisation scale, and $(-)^{n_f}$ is $+1$ for bosons and $-1$ for fermions, respectively. Since we are tracking the singlet scalar field for FOPT, we consider its VEV, denoted by $M$ as the renormalisation scale as $\mu = M = \langle S \rangle$. We denote the singlet scalar as $S=(\phi+M)/\sqrt{2}$. The relevant field dependent masses along with their dof are
\begin{equation}
    m_\eta^2=\lambda_7 \phi^2/2 \; (n_\eta = 4), m_s^2=3\lambda_s \phi^2 \; (n_s =1), m_{y_i}^2=2 y_i^2 \phi^2 \; (n_y=2)
\end{equation}

Thermal contributions to the effective potential are given by
\begin{align}
V_{\rm th} = \sum_i \left(\frac{n_{\rm B_i}}{2\pi^2}T^4 J_B \left[\frac{m_{\rm B_i}}{T}\right] - \frac{n_{\rm F_{i}}}{2\pi^2}T^4 J_F \left[\frac{m_{\rm F_{i}}}{T}\right]\right),
\end{align}
where $n_{B_i}$ and $n_{F_i}$ denote the dof of the bosonic and fermionic particles, respectively.
In this expressions, $J_B$ and $J_F$ functions are defined by following functions:
\begin{align}
&J_B(x) =\int^\infty_0 dz z^2 \log\left[1-e^{-\sqrt{z^2+x^2}}\right] \label{eq:J_B},\\
&J_F(x) =   \int^\infty_0 dz z^2 \log\left[1+e^{-\sqrt{z^2+x^2}}\right].
\end{align}
In the calculation of the thermal potential, we also consider the contribution from the daisy diagrams such that the total thermal potential reads $V_T (\phi, T) = V_{\rm th} + V_{\rm daisy} (\phi, T)$. This has to be done in order to improve the perturbative expansion during the FOPT \cite{Fendley:1987ef,Parwani:1991gq,Arnold:1992rz}. Such corrections can be implemented in two ways by inserting thermal masses into the zero-temperature field dependent masses. In one of these resummation prescriptions, known as the Parwani method \cite{Parwani:1991gq}, thermal corrected field dependent masses are used. In the other prescription, known as the Arnold-Espinosa method \cite{Arnold:1992rz}, the effect of the daisy diagram is included only for Matsubara zero-modes inside $J_B$ function defined above. In our work, we use the Arnold-Espinosa method. The thermal part of the potential, including the daisy contributions can now be written as
\begin{align}
    V_T(\phi,T) &= V_{\rm th} + V_{\rm daisy}(\phi,T), \\
    V_{\rm daisy}(\phi,T) &= -\sum_i \frac{g_i T}{12\pi}\left[ m^3_i(\phi,T) - m^3_i(\phi) \right], \nonumber
\end{align}
wherein, $V_{\rm th}$ is the thermal correction and $V_{\rm daisy}$ is the daisy subtraction \cite{Fendley:1987ef,Parwani:1991gq,Arnold:1992rz}. Denoting $m^2_i(\phi,T)=m^2_i(\phi) + \Pi_i(T)$, the relevant thermal masses can be written as \cite{Cline:2008hr}
\begin{align*}
\Pi_\eta(T)=(\frac{g_2^2}{8}+\frac{g_1^2+g_2^2}{16}+\frac{\lambda_2}{2}+\frac{\lambda_3+\lambda_4}{12})T^2 \\
\Pi_s(T)=(\frac{\lambda_s}{4}+\frac{\lambda_6}{3}+\frac{\lambda_7}{3}+\frac{y_1^2}{8}+\frac{y_2^2}{8}+\frac{y_2^2}{8})T^2.
\end{align*}

The FOPT proceeds via tunnelling, and the corresponding spherical symmetric field configurations known as bubbles are nucleated followed by expansion and coalescence. For recent reviews of FOPT in cosmological context, please refer to \cite{Mazumdar:2018dfl,Hindmarsh:2020hop}. The tunnelling rate per unit time per unit volume can be estimated as
\begin{align}
\Gamma (T) = \mathcal{A}(T) e^{-S_3(T)/T},
\end{align}
where $\mathcal{A}(T)\sim T^4$ and $S_3(T)$ are determined by the dimensional analysis and given by the classical configurations, called bounce, respectively. At finite temperature, the $O(3)$ symmetric bounce solution \cite{Linde:1980tt} can be obtained by solving the following equation
\begin{align}
    \frac{d^2 \phi}{dr^2}+\frac{2}{r}\frac{d\phi}{dr} = \frac{\partial V_{\rm tot}}{\partial \phi}\label{eq:bounce diff}.
\end{align}
The boundary conditions required to solve the above differential equation are
\begin{align}
\phi(r\to \infty)= \phi_{\rm false},~~~\left.\frac{d\phi}{dr}\right|_{r=0} =0,\label{eq:boundary condition}
\end{align}
where $\phi_{\rm false}$ denotes the position of the false vacuum. Using $\phi$ governed by the above equation and boundary conditions, the bounce action can be written as
\begin{align}
    S_3 =\int_0^{\infty} dr 4\pi r^2 \left[\frac{1}{2}\left(\frac{d\phi}{dr}\right)^2 +V_{\rm tot}(\phi,T)\right].
    \label{s3eq}
\end{align}
The temperature at which the bubbles are nucleated is called the nucleation temperature $T_n$. This can be calculated by comparing the tunnelling rate to the Hubble expansion rate as
\begin{align}
    \Gamma (T_n) = {\bf H}^4(T_n).
\end{align}
Here, assuming the usual radiation dominated universe, the Hubble parameter is given by ${\bf H}(T)\simeq 1.66\sqrt{g_*}T^2/M_{\rm Pl}$ with $g_*$ being the dof of the radiation component.
Thus, the rate comparison equation above leads to
\begin{align}
    \frac{S_3(T_n)}{T_n} \simeq 140, \label{eq:nucleation temperature}
\end{align}
for $g_*\sim 100$ and $T_n \sim 100$ GeV while for lower temperature near MeV where $g_*\sim 10$, the above ratio becomes larger. For higher nucleation temperature, as we have in the present scenario, the ratio $\frac{S_3(T_n)}{T_n}$ becomes smaller than the one quoted above. If $\phi(T_n) / T_n>1$ is satisfied, where $\phi(T_n)$ is the singlet scalar VEV at the nucleation temperature, $T=T_n$, the corresponding phase transition is conventionally called \textit{strong} first order. Alternatively, the ratio at critical temperature namely $\phi(T_c) / T_c \equiv v_c/T_c >1$ is also used as the strength of the FOPT. The critical temperature $T_c$ corresponds to the temperature where the two minima of the potential are degenerate.

\begin{table}
    \centering
    \begin{tabular}{|c|c|c|c|c|c|c|c|c|c|c|}
    \hline
       & $v_c$ (GeV) & $T_c$ (GeV) & v (GeV) & $v_c/T_c$ & $\lambda_7 (0)$ & $Y' (0)$ & $\lambda_s (0)$  \\
        \hline
       BP1 &  9540.78 & 2374 & 9901.77 & 4.01 & 1.5 & 0.5 & 0.02 \\
       \hline
       BP2 &  9599.59 & 2407 & 9968.24 & 3.98 & 1.6 & 0.4 & 0.02 \\
       \hline
       BP3 &  9734.10 & 2422 & 9997.27 & 4.01 & 1.2 & 0.2 & 0.02 \\
       \hline
       BP4 &  9648.57 & 2391 & 9988.00 & 4.03 & 1.4 & 0.3 & 0.02 \\
       \hline
    \end{tabular}
    \caption{Benchmark parameters of the model and other details of the FOPT in conformal scotogenic model.}
    \label{tab1}
\end{table}

In order to simplify the bounce calculation, we write the zero temperature one-loop effective potential as \cite{Jinno:2016knw, Iso:2009nw}
\begin{align}
     V_{0} &= V_{\rm tree} + V_{\rm CW}, \nonumber \\
    &= \frac{1}{4} \lambda_S(t) G^4(t) \phi^4
\end{align}
where $t={\rm log}(\phi/\mu)$ with $\mu=M$ being the scale of renormalisation. $G(t)$ is given by
\begin{align}
    G(t) = e^{-\int^t_0 dt' \gamma(t')},\; \gamma(t) = \frac{1}{32\pi^2} {\rm Tr}[Y'^{\dagger} Y'],
\end{align}
The Yukawa couplings and quartic coupling at the renormalisation scale are calculated by solving the renormalisation group evolution (RGE) equations of the model given in Appendix \ref{appen1}. Taking the renormalisation scale $\mu$ to be $M$, the condition $\frac{dV}{d\phi}|_{\phi=M}=0$ leads us to the relation,
\begin{equation}
    10\lambda^2_s(0) + 32 \pi^2 \lambda_s(0) +3 Y'^2(0) \lambda_s(0) -3Y'^4(0) + \lambda^2_6(0)+ \lambda^2_7(0)=0
    \label{minima1}
\end{equation}
assuming all three Yukawa couplings to be identical for simplicity. In order to get the required potential profile, the relative magnitude of the couplings $Y'$ and $\lambda_7$ are important as we will discuss below. The other quartic coupling $\lambda_6$ needs to be small in order to get the desired electroweak symmetry breaking at later stages.

Apart from finding the nucleation and critical temperature, it is also required to estimate the epoch when the FOPT gets completed. The corresponding temperature is known as the percolation temperature $T_p$, typically defined as the temperature at which significant volume of the universe is converted from the symmetric phase (false vacuum) to the broken phase (true vacuum). Adopting the prescription given in \cite{Ellis:2018mja, Ellis:2020nnr},
the percolation temperature $T_p$ is obtained from the probability of finding a point still in the false vacuum
given by
%
\begin{align}
    \mathcal{P}(T) = e^{-\mathcal{I}(T)}, \nonumber
\end{align}  
where
\begin{align}
    \mathcal{I}(T) = \frac{4\pi}{3}\int^{T_c}_T \frac{dT'}{T'^4}\frac{\Gamma(T')}{{\bf H}(T')}\left(\int^{T'}_T \frac{d\tilde{T}}{{\bf H}(\tilde{T})}\right)^3.
\end{align}
The percolation temperature is then calculated by using  $\mathcal{I}(T_p) = 0.34$ \cite{Ellis:2018mja} (implying that at least $34\%$ of the comoving volume is occupied by the true vacuum).

\begin{figure}
    \centering
    \includegraphics[scale=0.5]{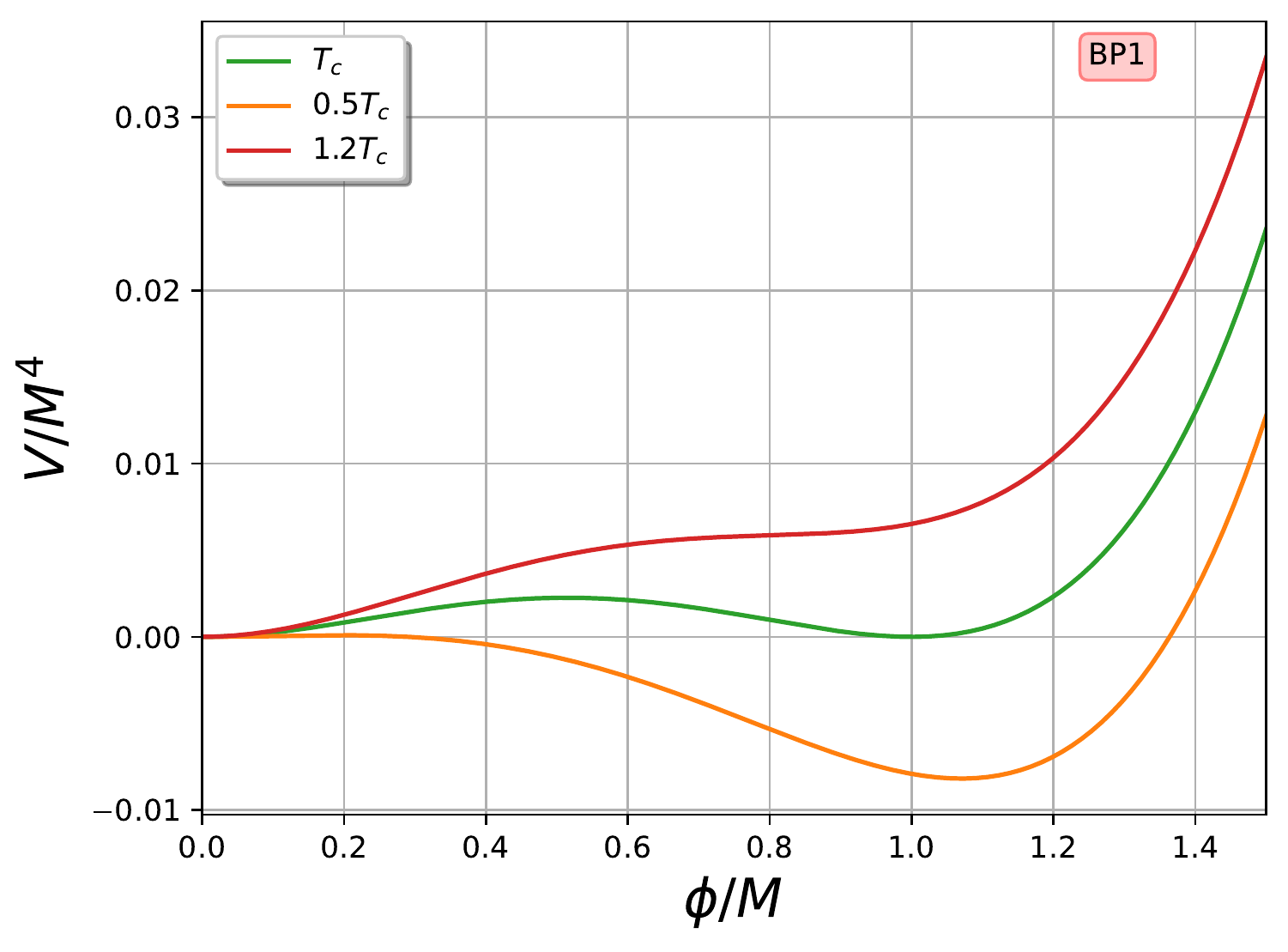}
        \includegraphics[scale=0.5]{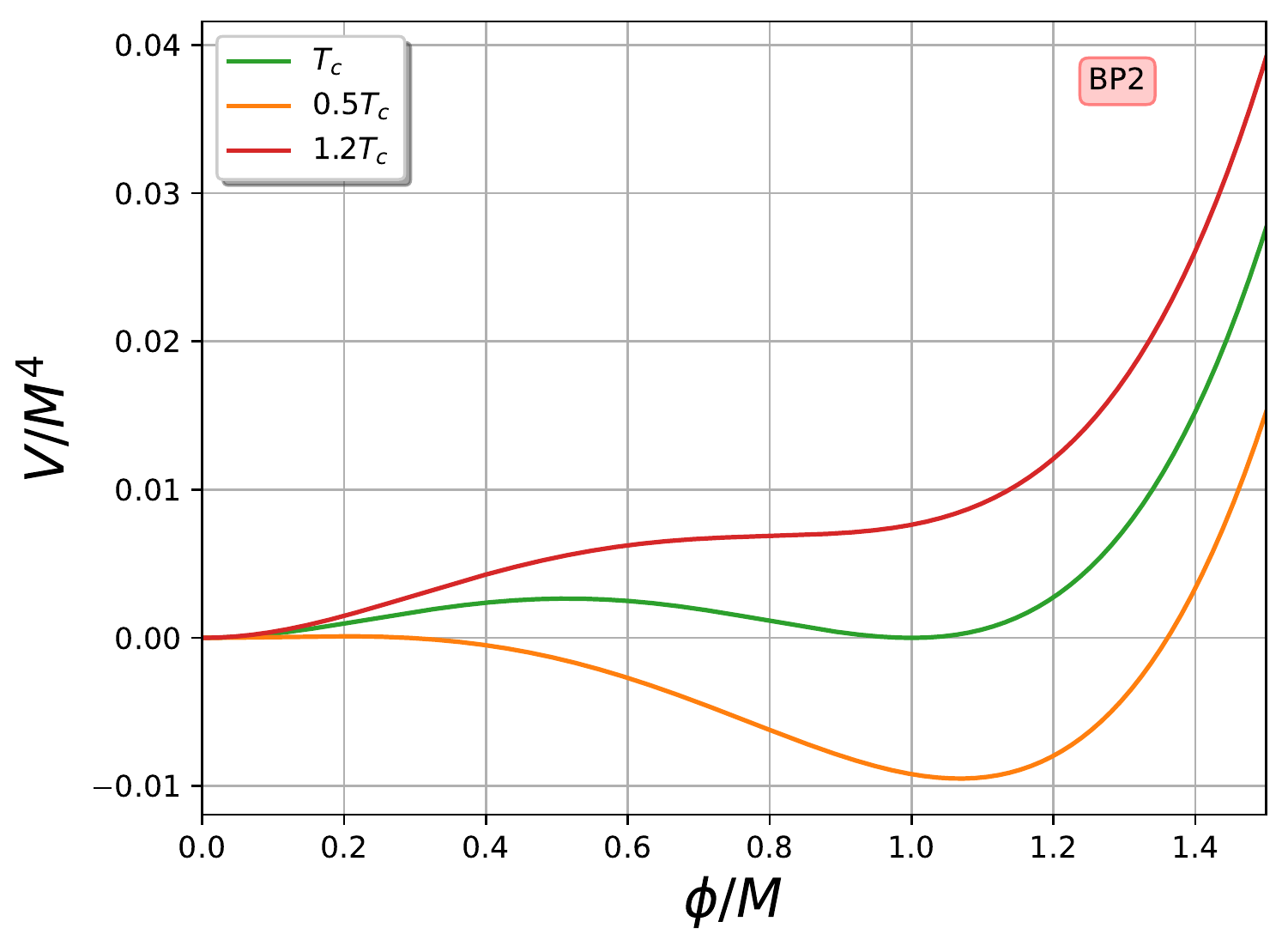} \\
            \includegraphics[scale=0.5]{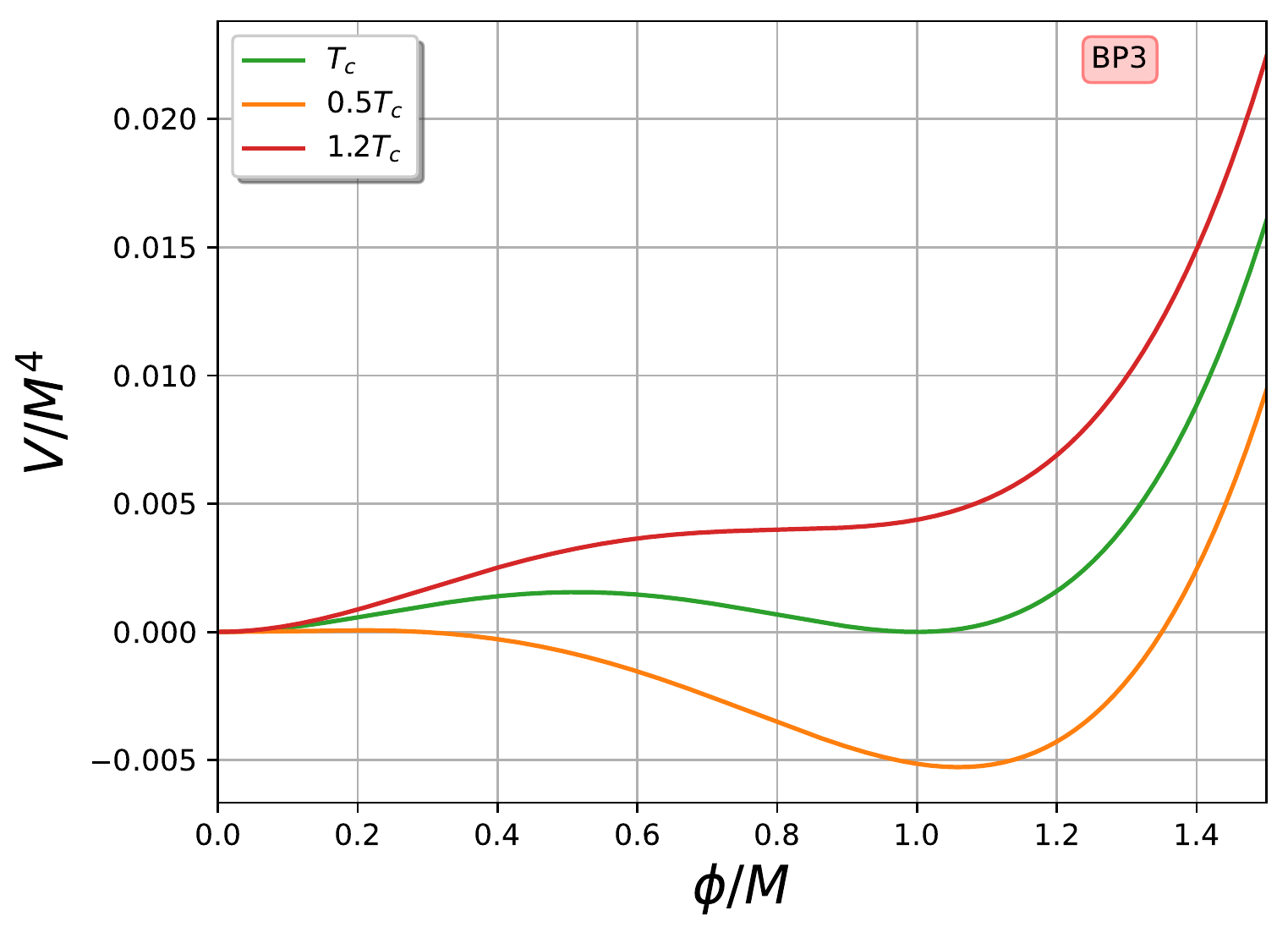}
                \includegraphics[scale=0.5]{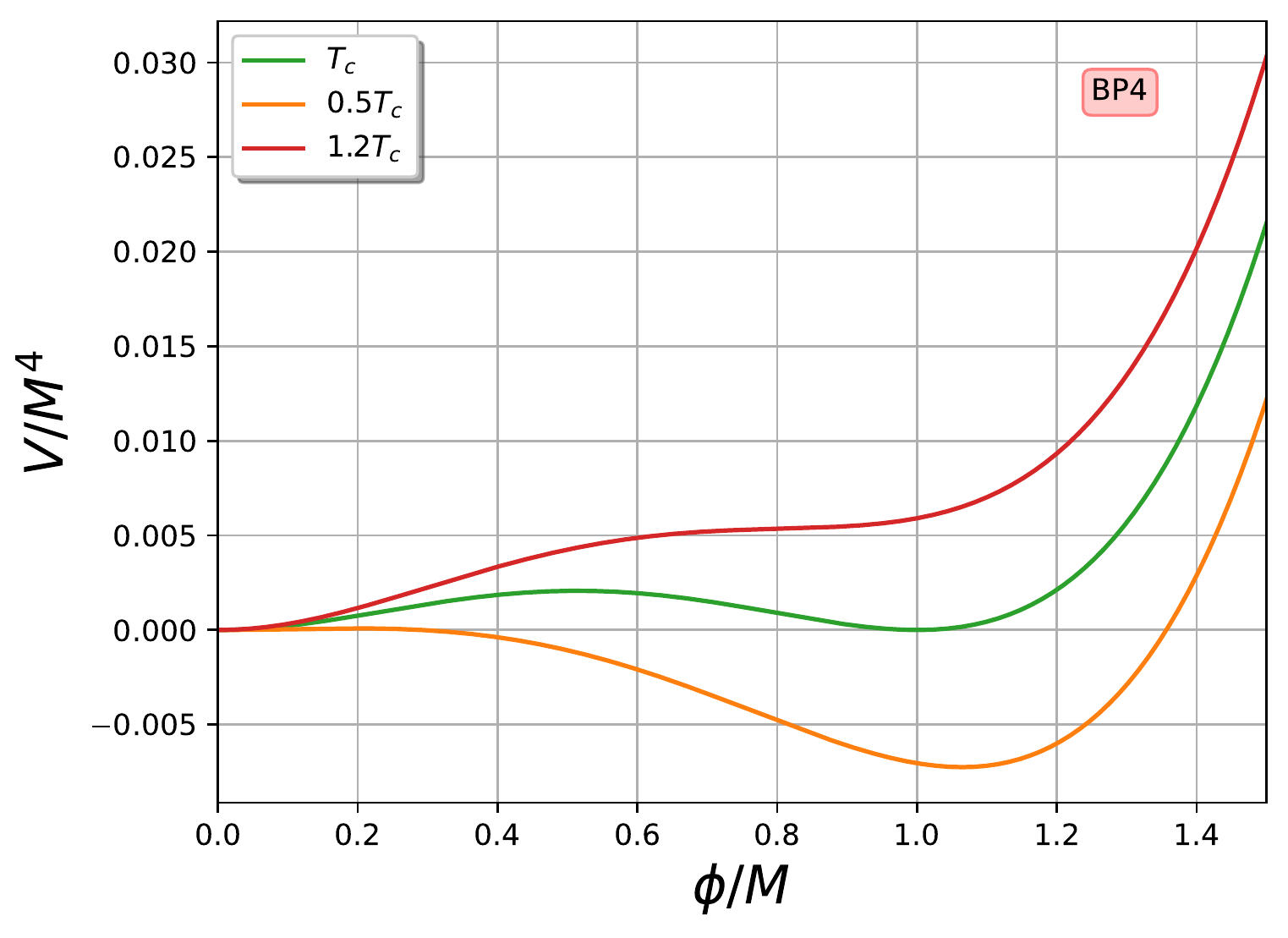}
                \caption{Shape of the potential at, above and below the critical temperature $T_c$ for chosen benchmark points shown in Table \ref{tab1}.}
    \label{fig1}
\end{figure}

\begin{figure}
    \centering
    \includegraphics{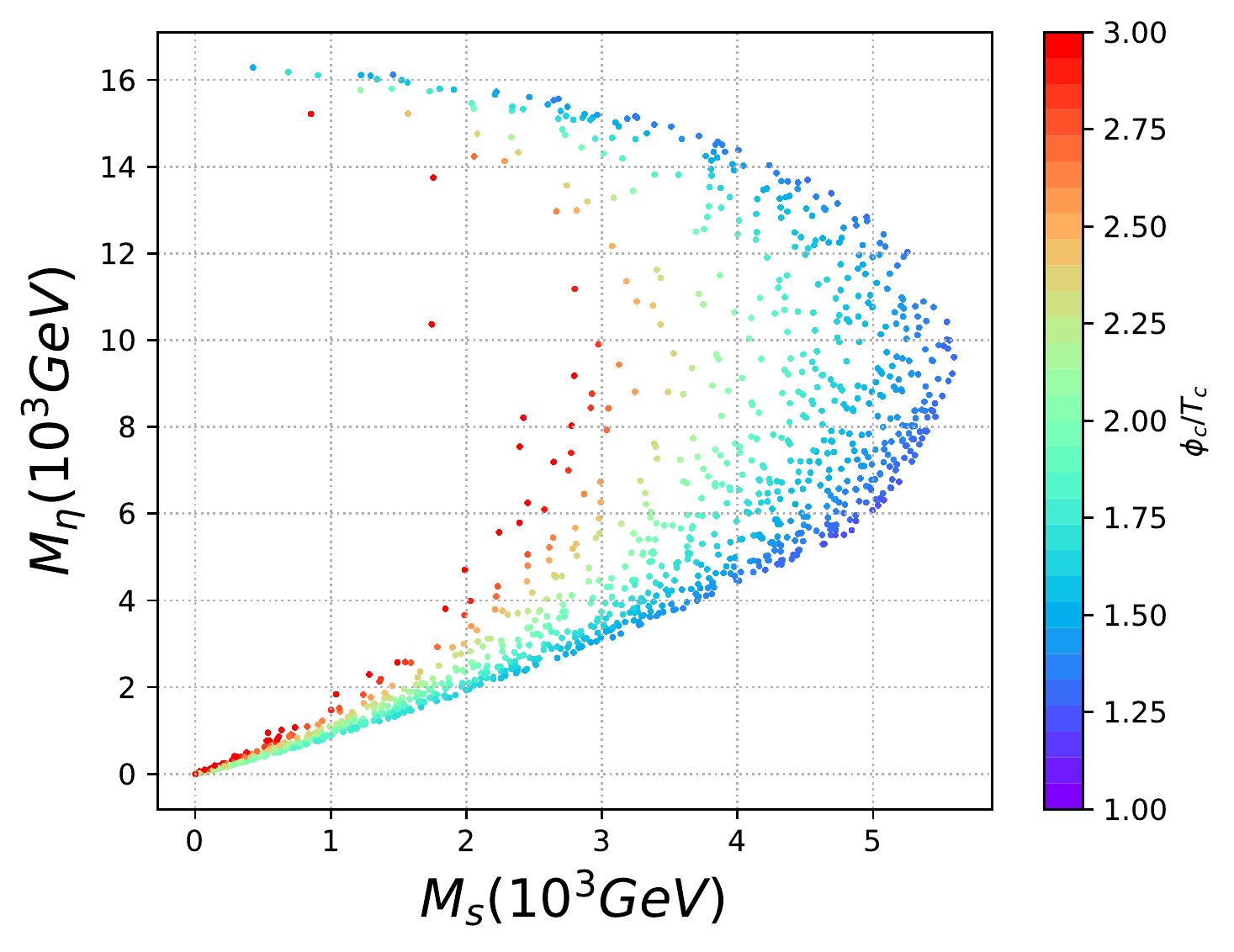}
    \caption{Parameter space in inert doublet Mass $M_\eta$ versus scalar singlet mass $M_S$ plane consistent with a FOPT in conformal scotogenic model. The colour code indicates the strength of the FOPT.}
    \label{fig2}
\end{figure}

We then implement the model in \texttt{PhaseTracer} \cite{Athron:2020sbe} to find the parameter space consistent with a FOPT. For a few benchmark points given in table \ref{tab1}, we show the potential profile in Fig. \ref{fig1}. For all these benchmarks, one can clearly see a barrier between the two minima at the critical temperature, indicating a FOPT. Such degenerate minima lead to the formation of bubbles which subsequently produce gravitational waves. We also perform a numerical scan to show the model parameter space in new scalar masses in Fig. \ref{fig2}. As shown in the colour code, the strength of the FOPT can be large $v_c/T_c \geq 3$ for certain region of the parameter space. This, along with the supercooled nature of the FOPT helps in enhancing the strength of the resulting gravitational waves emitted, as we discuss below.

\section{Stochastic Gravitational Waves from FOPT}
\label{sec4}
A FOPT can lead to the formation of stochastic gravitational waves (GW) background primarily due to three distinct mechanisms: the bubble collisions~\cite{Turner:1990rc,Kosowsky:1991ua,Kosowsky:1992rz,Kosowsky:1992vn,Turner:1992tz}, the sound wave of the plasma~\cite{Hindmarsh:2013xza,Giblin:2014qia,Hindmarsh:2015qta,Hindmarsh:2017gnf} and the turbulence of the plasma~\cite{Kamionkowski:1993fg,Kosowsky:2001xp,Caprini:2006jb,Gogoberidze:2007an,Caprini:2009yp,Niksa:2018ofa}. The amplitude of such GW signal crucially depends upon two quantities: the amount of vacuum energy (or latent heat) released during the transition as well as the duration of the transition.

In order to calculate the energy released during the FOPT, we first find the free energy difference between the true and the false vacuum as
\begin{equation}
    \Delta V_{\rm tot} \equiv V_{\rm tot}(\phi_{\rm false},T)- V_{\rm tot}(\phi_{\rm true},T).
\end{equation}
As a result of the bubble nucleation, the amount of vacuum energy released during the FOPT, in the units of radiation energy density of the universe, $\rho_{\rm rad}= g_*\pi^2 T^4/30 $, is given by
\begin{align}
    \alpha_* =\frac{\epsilon_*}{\rho_{\rm rad}},
    \label{alphastar}
\end{align}
with
\begin{align}
    \epsilon_* = \left[\Delta V_{\rm tot} - \frac{T}{4} \frac{\partial \Delta V_{\rm tot}}{\partial T}\right]_{T=T_*},
\end{align}
which is also related to the change in the trace of the energy-momentum tensor across the bubble wall \cite{Caprini:2019egz,Borah:2020wut}.

On the other hand, the duration of the FOPT, denoted by the parameter $\beta$, is defined as \cite{Caprini:2015zlo}
\begin{align}
\frac{\beta}{{\bf H}(T)} \simeq T\frac{d}{dT} \left(\frac{S_3}{T} \right).
\end{align}
Here, $\alpha_*$ and $\beta/{\bf H}(T)$ are evaluated at the nucleation temperature $T=T_*$ with $S_3$ being evaluated using Eq. \eqref{s3eq}. In order to calculate the action numerically, we use a fit for the actual potential which matches very well with the actual potential, as shown in Appendix \ref{appen2}. For the benchmark points discussed earlier, we calculate these key parameters and show them in Table \ref{tab2} along with other relevant parameters.

\begin{table}
    \centering
    \begin{tabular}{|c|c|c|c|c|c|c|c|c|}
    \hline
       & $T_c$ (GeV) & $v_c/T_c$ & $T_n$ (GeV) & $T_p$ (GeV) & $(\beta/{\bf H}_*)$ & $v_J$ & $\alpha_*$\\
        \hline
       BP1 &  2374 &  4.01 & 810.86 & 808.35 & 77.43 & 0.93 & 0.99\\
       \hline
       BP2 & 2407 &  3.98 & 1098.88 & 1084.30 & 168.18 & 0.86 & 0.35\\
       \hline
       BP3 &  2422 &  4.01 & 834.79 & 823.52 & 125.87 & 0.91 & 0.61\\
       \hline
       BP4 &  2391 & 4.03 & 975.59 & 962.33 & 150.66 & 0.88 & 0.44\\
       \hline
    \end{tabular}
    \caption{Benchmark parameters consistent with FOPT in conformal scotogenic model, along with the FOPT related parameters calculated for GW spectrum estimation.}
    \label{tab2}
\end{table}

\begin{figure}
    \centering
    \includegraphics[scale=0.40]{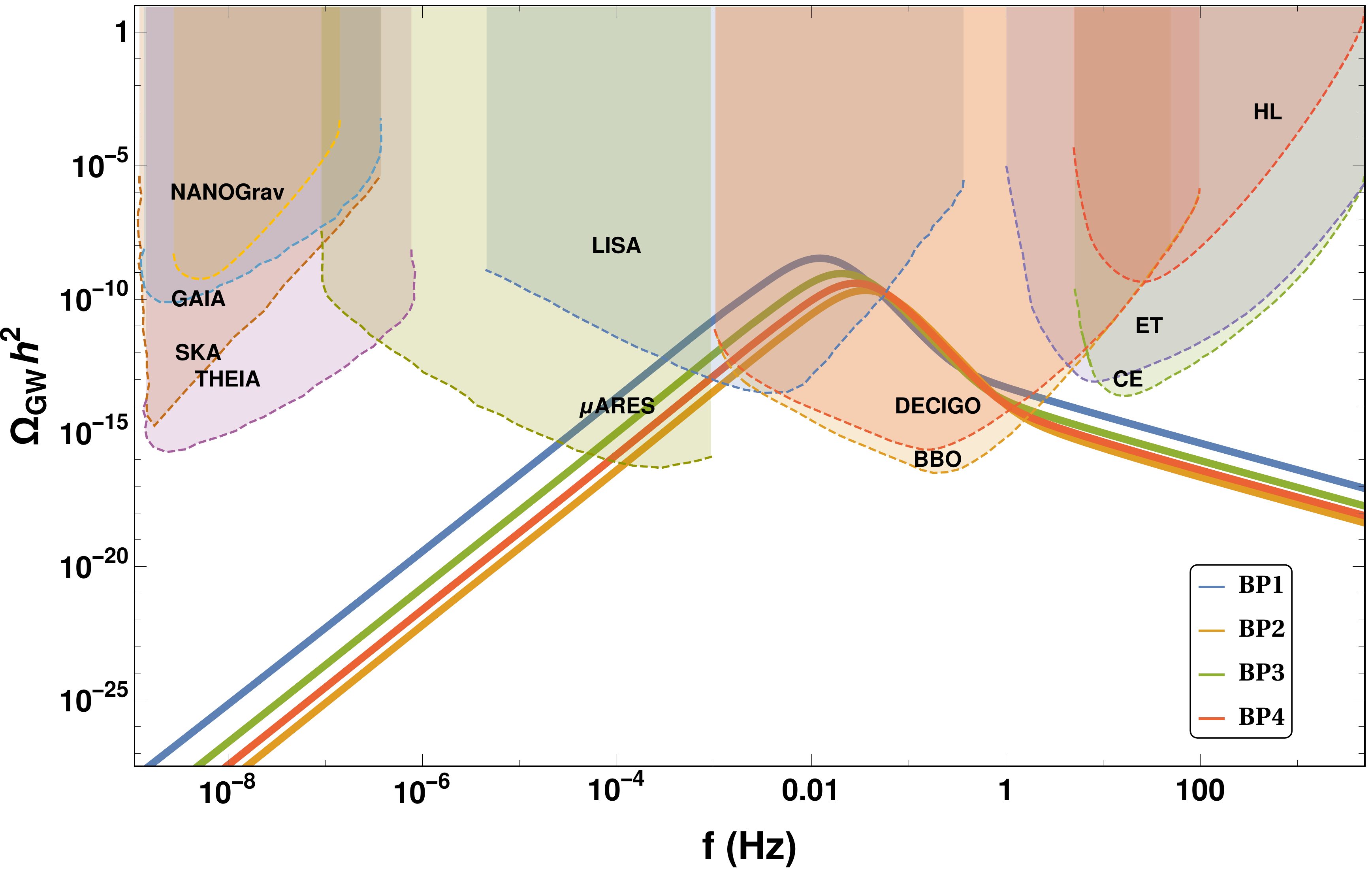}
    \caption{Gravitational wave spectrum from FOPT in conformal scotogenic model for four different benchmark points given in table \ref{tab2}. Different coloured curves show the sensitivities from GW search experiments like LISA, BBO, DECIGO, HL (aLIGO), ET, CE, NANOGrav, SKA, GAIA, THEIA and $\mu$ARES.}
    \label{fig3}
\end{figure}

Now, considering the three contributions to GW production mentioned above, the corresponding GW power spectrum can be written as \cite{Cai:2017tmh}
\begin{align}
    \Omega_{\rm GW}(f) &= \Omega_\phi(f) + \Omega_{\rm sw}(f) + \Omega_{\rm turb}(f).
\end{align}
In general, each of these contributions can be characterised by its own peak frequency and each GW spectrum can be written in parametric form as
\begin{align}
    h^2\Omega(f) &= \mathcal{R}\Delta(v_w)\left(\frac{\kappa \alpha_*}{1+\alpha_*}\right)^p\left(\frac{{\bf H_*}}{\beta}\right)^q\mathcal{S}(f/f_{\rm peak}).
\end{align}
Here, the pre-factor $\mathcal{R}\simeq 7.69\times 10^{-5}g^{-1/3}_*$ takes in account the red-shift of the GW energy density, $\mathcal{S}(f/f_{\rm peak})$ parametrises the shape of the spectrum and $\Delta(v_w)$ is the normalization factor which depends on the bubble wall velocity $v_w$. The Hubble parameter at the nucleation temperature $T=T_n$ is denoted by ${\bf H_*}$. For bubble collision as source, the spectrum can be written as \cite{Cai:2017tmh}
\begin{equation}
    \Omega_\phi h^2 = 1.67 \times 10^{-5} \left ( \frac{100}{g_*} \right)^{1/3} \left(\frac{{\bf H_*}}{\beta}\right)^2 \left(\frac{\kappa \alpha_*}{1+\alpha_*}\right)^2 \frac{0.11 v^3_w}{0.42+v^2_w} \frac{3.8(f/f_{\rm peak})^{2.8}}{1+2.8 (f/f_{\rm peak})^{3.8}}
\end{equation}
with the peak frequency being given by 
\begin{equation}
    f_{\rm peak} = 1.65 \times 10^{-5} {\rm Hz} \left ( \frac{g_*}{100} \right)^{1/6} \left ( \frac{T_n}{100 \; {\rm GeV}} \right ) \frac{0.62}{1.8-0.1v_w+v^2_w} \left(\frac{\beta}{{\bf H_*}}\right).
\end{equation}
Similarly, the other contributions can also be written following \cite{Cai:2017tmh} and references therein.    In order to calculate the bubble wall velocity., we first calculate the Jouguet velocity~\cite{Kamionkowski:1993fg, Steinhardt:1981ct, Espinosa:2010hh}\footnote{Also see Refs.~\cite{Huber:2013kj,Leitao:2014pda,Dorsch:2018pat,Cline:2020jre}, for the discussion of the bubble wall velocity $v_w$.}:
\begin{align}
    v_J = \frac{1/\sqrt{3} + \sqrt{\alpha^2_* + 2\alpha_*/3}}{1+\alpha_*}.
\end{align}
The bubble wall velocity is then calculated as \cite{Lewicki:2021pgr}
\begin{equation}
    v_w = 
    \begin{cases}
    \sqrt{\frac{\Delta V_{\rm tot}}{\alpha_* \rho_{\rm rad}}} & \text{if} \,\, \sqrt{\frac{\Delta V_{\rm tot}}{\alpha_* \rho_{\rm rad}}} < v_J \\
    1 & \text{if}  \,\, \sqrt{\frac{\Delta V_{\rm tot}}{\alpha_* \rho_{\rm rad}}} \geq v_J. \\
    \end{cases}
\end{equation}
The total GW spectrum after summing over the contributions from all three sources is shown in Fig. \ref{fig3} for the benchmark points shown in Table \ref{tab2}. The experimental sensitivities of NANOGrav \cite{McLaughlin:2013ira}, SKA \cite{Weltman:2018zrl}, GAIA \cite{Garcia-Bellido:2021zgu}, THEIA \cite{Garcia-Bellido:2021zgu}, $\mu$ARES \cite{Sesana:2019vho}, LISA\,\cite{AmaroSeoane2012LaserIS}, DECIGO \cite{Kawamura:2006up}, BBO\,\cite{Yagi:2011wg}, ET\,\cite{Punturo_2010}, CE\,\cite{LIGOScientific:2016wof} and aLIGO \cite{LIGOScientific:2014pky} are shown as shaded regions of different colours. Since the FOPT is occurring at a scale above the electroweak scale, the peak frequencies as well as the amplitudes are around the LISA sensitivity and hence remain verifiable in near future.

\begin{figure}
    \centering
    \includegraphics[scale=0.75]{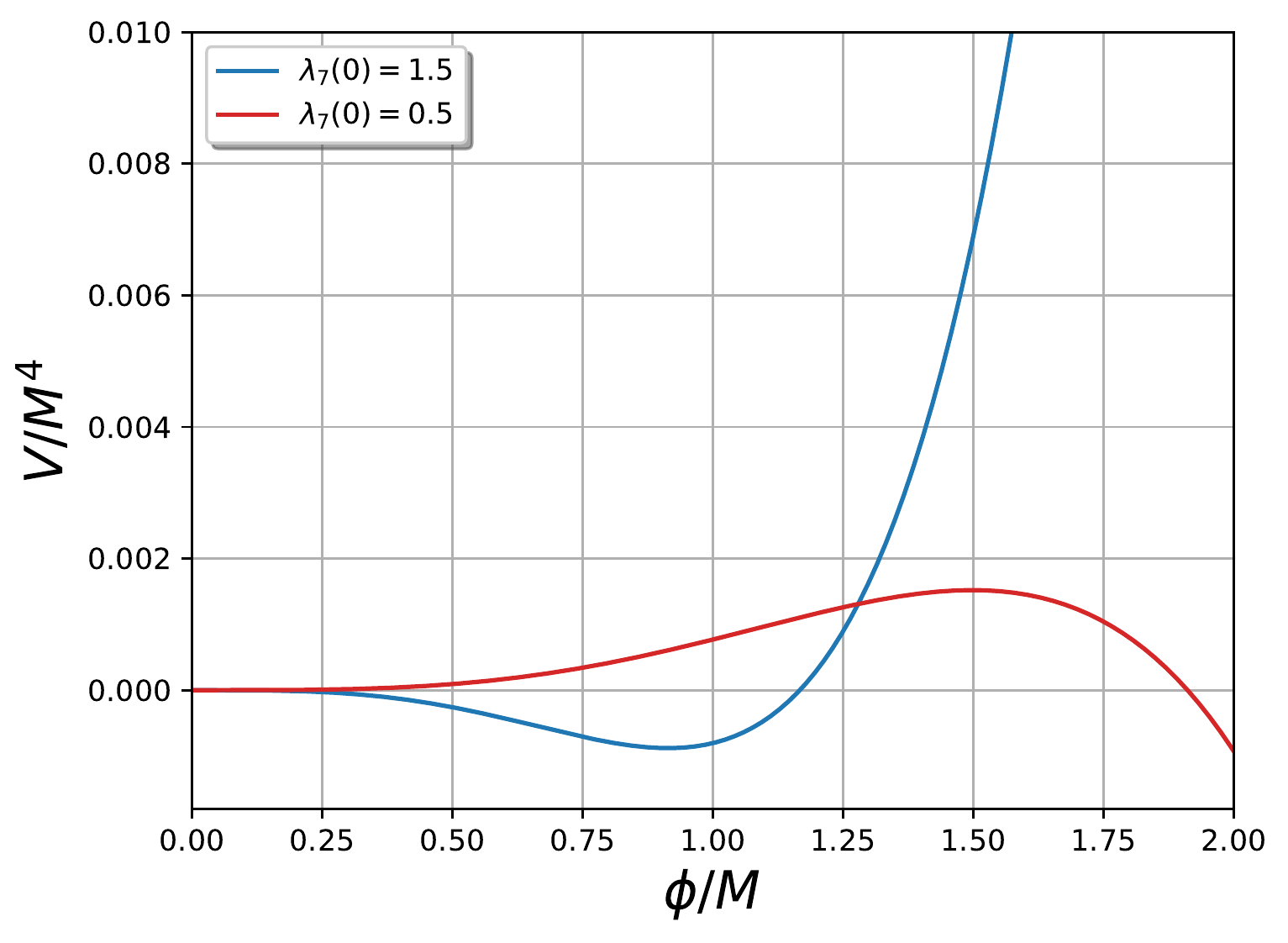}
    \caption{Zero temperature effective potential at one-loop for two different values of singlet-$\eta$ coupling $\lambda_7$, considering the singlet-RHN Yukawa coupling $Y'(0)=0.8$.}
    \label{fig4}
\end{figure}

\section{Dark Matter and Leptogenesis via FOPT}
\label{sec5}
In order to realise leptogenesis from decay, one needs to ensure that at least one of the RHNs remain heavier than the scalar doublet $\eta$. Since both $\eta$ and RHNs acquire masses during the FOPT, this helps in realising DM and leptogenesis simultaneously. However, the desired profile of the scalar potential of singlet scalar $S$ as well as the minimisation condition given in Eq. \eqref{minima1} pose a problem. As understood from the FOPT, the scalar singlet potential has one unique minima $\phi=0$ at very high temperature $T \gg M$ with the effective self-quartic coupling $\lambda_s >0$. However, for low temperature $T \ll M$, the self quartic coupling turns negative and $\phi=0$ should become a false vacuum. This is however, not possible unless we have $\lambda^2_7 (0) > 3 Y'^2 (0) $. This can be seen from Fig. \ref{fig4} where the zero-temperature effective potential is shown for two different values of $\lambda_7(0)$ while keeping singlet-RHN Yukawa fixed $Y'(0)=0.8$. Clearly, for smaller $\lambda_7(0)$, we can not achieve the desired potential profile at zero temperature. 

In order to circumvent this problem, we consider a hybrid of scotogenic and type I seesaw model without increasing the number of fields. Out of the three RHNs in conformal scotogenic model, we consider two of them to be $Z_2$ even such that they couple to the SM lepton doublets via usual Higgs doublet as $y_D \overline{L} \tilde{\Phi_1} N_{2,3}$. The other RHN namely, $N_1$ is $Z_2$-odd and couple to the SM lepton doublets via $\eta$ as before. Thus, two of the active neutrinos will receive non-zero mass from type I seesaw while the third one will receive scotogenic contribution at one-loop. The scalar potential as well as singlet scalar coupling to RHNs remain same as before and hence we still require $\eta$ to be heavier than the RHNs. Therefore $N_1$ is our DM candidate and $N_{2,3}$ can decay into $L \Phi_1$ to generate the required lepton asymmetry.

\begin{figure}
    \centering
    \includegraphics[scale=0.40]{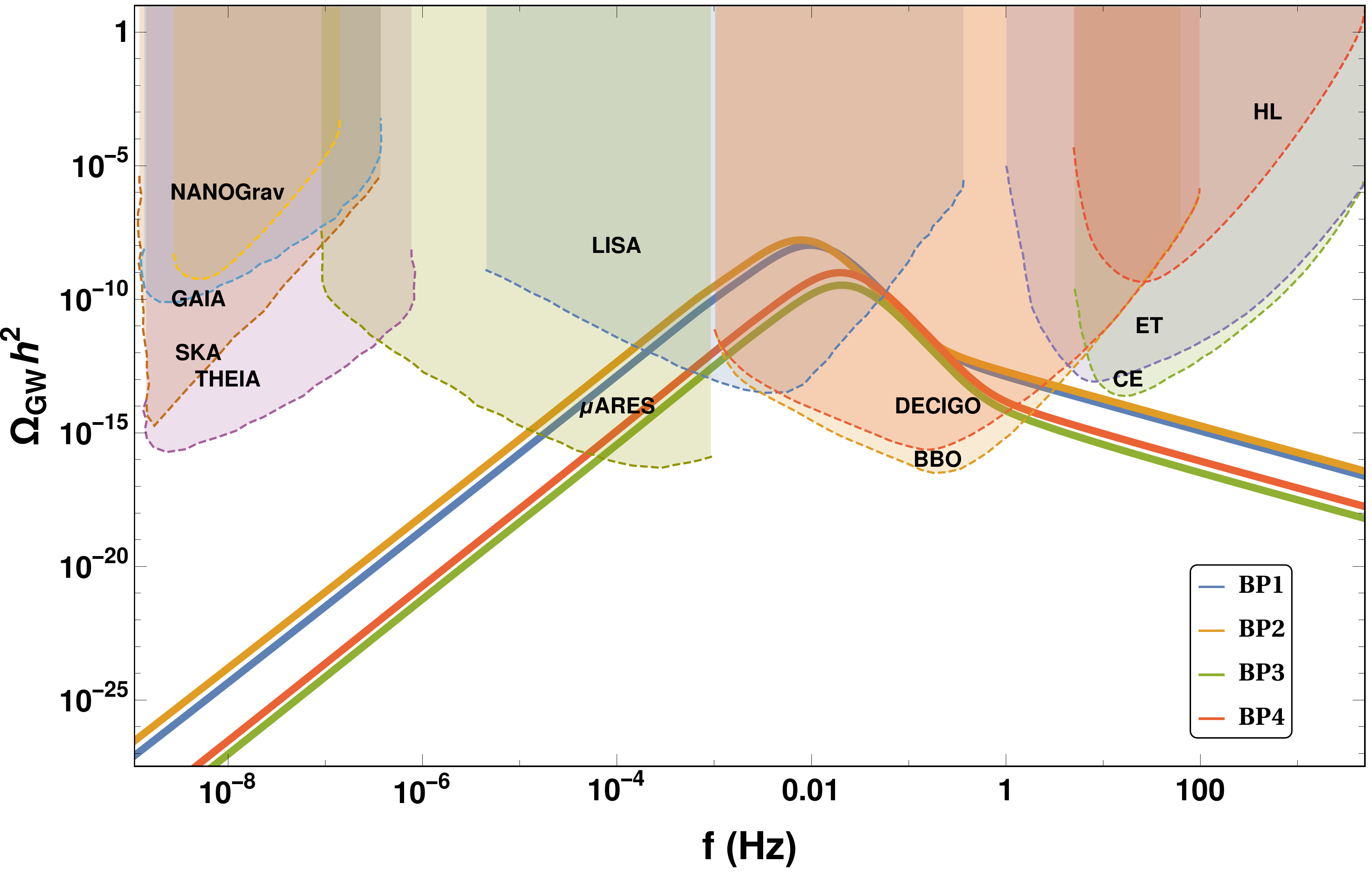}
    \caption{Gravitational wave spectrum from FOPT in the hybrid model for four different benchmark points given in table \ref{tab3}. Different coloured curves show the sensitivities from GW search experiments like LISA, BBO, DECIGO, HL (aLIGO), ET, CE, NANOGrav, SKA, GAIA, THEIA and $\mu$ARES.}
    \label{fig5}
\end{figure}

We first identify a few benchmark points consistent with the FOPT and the mass hierarchy among RHN and scalar doublet $\eta$ required to have successful leptogenesis and DM phenomenology. The benchmark points along with other details calculated for the GW spectrum are shown in table \ref{tab3}. The corresponding GW spectrum is shown in Fig. \ref{fig5}. Clearly, choosing one RHN lighter and making the heavier RHNs $Z_2$ even does not change the FOPT details significantly and hence we obtain similar benchmark parameters and GW spectrum like before. 

\begin{table}
    \centering
    \begin{tabular}{|c|c|c|c|c|c|c|c|c|c|c|c|c|}
    \hline
       & $v_c$  & $T_c$  & v & $v_c/T_c$ & $\lambda_7 (0)$  & $Y'_2 (0)$ &  $\lambda_s (0)$ & $T_n$ & $T_p$ & $(\beta/{\bf H}_*)$ & $v_J$ &  $\alpha_*$\\
       & (GeV) & (GeV) & (GeV) &  & & $\approx Y'_3 (0)$ & & (GeV) & (GeV) & & & \\
        \hline
       BP1 &  9634.17 & 2521 & 9934.17 & 3.82 & 1.5 & 0.5 & 0.02 & 988.32 & 974.98 & 151.06 & 0.89 & 0.46 \\
       \hline
       BP2 &  9553.88 & 2416 & 9757.60 & 3.95 & 1.6  & 0.7 & 0.02 & 896.89 & 887.67 & 110.66 & 0.91 & 0.66 \\
       \hline
       BP3 &  9698.63 & 2370 & 9988.01 & 4.09 & 1.2 &  0.3 & 0.02 & 779.49 & 770.72 & 103.96 & 0.92 & 0.80\\
       \hline
       BP4 &  9692.84 & 2391 & 9978.08 & 4.05 & 1.3 &  0.4 & 0.02 & 1207.09 & 1190.58  & 204.51 & 0.84 & 0.24\\
       \hline
    \end{tabular}
    \caption{Benchmark parameters and other details involved in the GW spectrum calculation of the hybrid model.}
    \label{tab3}
\end{table}

Now we implement the baryogenesis via relativistic bubble wall mechanism proposed in \cite{Baldes:2021vyz} to achieve leptogenesis in the hybrid model mentioned above. A first order phase transition in the singlet (S) sector will create bubbles such that the particles like $N_i, \eta$ entering the bubble will become massive due to $\langle S \rangle \neq 0$ inside the bubble. This is followed by $N_{2,3}$ decays into leptons and SM Higgs creating the leptonic asymmetry. On the other hand $N_1$ being lighter than $\eta$ become stable due to unbroken $Z_2$ symmetry and hence act like a DM candidate.

\begin{table}[]
    \centering
    \begin{tabular}{|c|c|c|c|c|c|c|}
       \hline & $\epsilon_N$  & $T_{\rm RH}$ (GeV) & $T_n$ (GeV)&  $M_{N_{2}} \approx M_{N_3}$ (GeV)& $y_D$ & $\Delta V_{\rm tot} $ (GeV)$^4$\\
        \hline 
         BP1 & $6.22\times10^{-8}$   & 988.32  & 988.32 &  4966.67 & $5.12\times10^{-8}$ & $1.75464\times 10^{13}$\\
        BP2 & $6.22\times 10^{-8}$  & 896.89 & 896.89 & 6826.16 & $6\times10^{-8}$ & $1.69737\times 10^{13}$\\
        BP3 & $6.22\times 10^{-8}$ & 779.49 & 779.49 & 2996.39 & $3.98\times10^{-8}$  &$1.15697\times 10^{13}$\\
    BP4 &  $6.22\times 10^{-8}$   & 1207.58 & 1207.58 & 3991.16 & $4.59\times 10^{-8}$ & $2.15494\times 10^{13}$\\
        \hline
    \end{tabular}
    \caption{CP asymmetry and other relevant details involved in leptogenesis calculation for the hybrid model.}
    \label{tab:asym}
\end{table}

For the leptogenesis we closely follow Ref.~\cite{Baldes:2021vyz,Dasgupta:2022isg} i.e., the {\it mass -gain} mechanism. Let us briefly mention the {\it mass-gain} mechanism employed in our work. Firstly, we need to ensure that the Lorentz boost of the bubble wall should be more than the Lorentz factor of the particle in the plasma frame 
\begin{align}
    \gamma_w >\gamma_N \sim \frac{M_N}{T_n}
    \label{eq:gammaw}
\end{align}
where $T_n$ is the nucleation temperature and $M_N = y_N v$ is the mass of the RHN coupling to the singlet scalar. Now, the above condition \eqref{eq:gammaw} pushes the RHN into the bubble while maintaining the equilibrium co-moving number density.
\begin{align}
    Y_N &= \frac{135}{8\pi^4}\xi(3)\frac{g_N}{g_*}
\end{align}
where $g_N$ and $g_*$ are the degrees of freedom of RHN $N$ and the total relativistic degrees of freedom in the energy density of the universe, respectively.

The final baryonic asymmetry is then written as follows 
\begin{align}
    Y_B &= \epsilon_N \kappa_{\rm sph}Y_N\left(\frac{T_n}{T_{RH}}\right)^3.
    \label{eq:eps}
\end{align}
where $\epsilon_N \simeq \sin(2\delta)/(16\pi)$ \cite{Pilaftsis:1998pd, Pilaftsis:2003gt} is the CP-asymmetry and $\delta$ is the relative CP phase between the RHNs (for resonant regime), $\kappa_{\rm Sph} = 8/23$ is the sphaleron conversion factor in the presence of two Higgs doublets~\cite{Kuzmin:1985mm}, and $T_{\rm RH}$ is the reheating temperature after the FOPT. $T_{\rm RH}$ is defined as $T_{\rm RH} = {\rm Max}[T_n, T_{\rm inf}]$ \cite{Baldes:2021vyz} where $T_{\rm inf}$ can be obtained from the following relation
\begin{equation}
    \frac{g_*\pi^2}{30}T^4_{\rm inf} = \Delta V_{\rm tot}.
\end{equation}
Using $\Delta V_{\rm tot}$ for the benchmark parameters given in table \ref{tab:asym}, we can calculate the corresponding $T_{\rm inf}$ and hence $T_{\rm RH}$. The $Y_B$ obtained in Eq.~\eqref{eq:eps} should then be compared with the observed baryon asymmetry normalized over the entropy density: $Y_B^{\rm obs}=(8.61\pm 0.05)\times 10^{-11}$~\cite{Planck:2018vyg}.

The above asymmetry is feasible after satisfying two condition
\begin{enumerate}
    \item The feasibility of decay $N_{2,3}\rightarrow LH$
    \item The wash-out from the dominant inverse decay to be suppressed.
\end{enumerate}
For the first condition we will need to consider the thermally corrected masses for the SM Higgs and lepton doublets at the reheating temperature~\cite{Giudice:2003jh}
\begin{align}
    M_H^2(T) &= \left(\frac{3}{16}g^2_2 + \frac{1}{16}g^2_1 + \frac{1}{4}y^2_t\right)T^2 \, , \nonumber \\
    M^2_L(T) &= \left(\frac{3}{32}g^2_2 + \frac{1}{32}g^2_1\right) T^2  \, , 
\end{align}
where $g_1$ and $g_2$ are the $U(1)_Y$ and $SU(2)_L$ gauge couplings respectively, and $y_t$ is the top quark Yukawa coupling. Therefore, at the reheating temperature after considering the coupling values at the electroweak scale\footnote{It should be noted that the values of these couplings do not change much between the electroweak scale and the reheating temperature for (multi) TeV-scale symmetry breaking considered here.} we get 
\begin{align}
    M_H(T_{\rm RH}) + M_L(T_{\rm RH})&\simeq 0.77 T_{\rm RH} , 
    \label{eq:therm1}
\end{align}
Hence for the feasibility of the decay we need the mass of RHN at the reheating temperature to be $M_N/T_{\rm RH}\gtrsim 0.77$.

As for the second condition we have taken the Dirac Yukawa coupling $y_D$, which is responsible for the wash-out, to be parameterized by the Casas-Ibarra parameterization~\cite{Casas:2001sr} for type I seesaw with two RHNs given by
\begin{align}
    Y_D &= \Lambda^{-1/2}\mathcal{O}\widehat{m}_{\nu}^{1/2}U^\dagger_{\rm PMNS} \, ,
    \end{align}
where $\Lambda=v^2_{\rm ew}/M_N$, ${\cal O}$ is an arbitrary complex orthogonal matrix, $\widehat{m}_\nu$ is the diagonal light neutrino mass matrix and $U_{\rm PMNS}$ is the light neutrino mixing matrix. Since only two RHNs contribute to type I seesaw, we consider the lightest active neutrino mass to be vanishing. Using the best-fit values of the light neutrino oscillation data~\cite{Gonzalez-Garcia:2021dve} for normal hierarchy and assuming ${\cal O}$ to be the identity matrix, we obtain  
    \begin{align}
    y_D \equiv \sum_\alpha y_{D_{1\alpha}} \sim  2.3\times 10^{-8}\left(\frac{M_N}{1 ~{\rm TeV} }\right)^{1/2} \, .
\end{align}
And proceeding with the above Dirac Yukawa we need to satisfy the following condition~\cite{Baldes:2021vyz} 
\begin{align}
    \frac{M_N}{T_{\rm RH}} \gtrsim \ln\left[\frac{y_D^2M_{\rm Pl}}{24\pi T_{\rm RH}}\left(\frac{M_N}{T_{\rm RH}}\right)^{5/2}\right], 
\end{align}
ensuring the inverse decay width to be suppressed. We calculate the required CP asymmetry and the Dirac Yukawa couplings for the four benchmark points and quote them in table \ref{tab:asym}. All these benchmark points satisfy the required baryon asymmetry due to the appropriate choice of Yukawa couplings and CP phase. As can be seen from the smallness of the Dirac Yukawa couplings, the decay width of the RHN remains small, also required from the resonant leptogenesis condition $M_3-M_2 \sim \Gamma_2/2$. This also justifies the semi-degenerate nature of RHNs $N_{2,3}$ in the benchmark choice of parameters. We also check that the benchmark points satisfy the above mentioned conditions to ensure the viability of the leptogenesis scenario we are implementing. 

For the standard vanilla leptogenesis scenario where  the \emph{massive} RHNs are in equilibrium, the baryon asymmetry in the weak washout regime $\Gamma/{\bf H}(T=M_N) \ll 1$ can be written as
\begin{align}
    Y_B &= \epsilon_N \kappa_{\rm sph}Y^{\rm eq}_N(T=M_N) ; \\
    Y^{\rm eq}_N(T)&= \frac{45}{4\pi^4}\frac{g_N}{g_*}\left(\frac{M_N}{T}\right)^2K_2(M_N/T),
    \label{eq:eps_std}
\end{align}
with $K_2$ being the modified Bessel function of 2nd order. Comparing with Eq. \eqref{eq:eps}, we can see that in FOPT scenario there arises an extra dilution factor $(T_n/T_{\rm RH})^3$ compared to the standard case. This was also noticed in earlier works \cite{Huang:2022vkf, Dasgupta:2022isg}. However, due to a different structure of our model in the absence of any additional gauge symmetry, we have $T_{\rm inf} < T_n$ resulting in $T_{\rm RH}=T_n$. Therefore, the final baryon asymmetry in our setup remains same as the standard one, but with the added advantage of detection prospects via stochastic GW observations. In other words, the model without FOPT is also consistent with successful leptogenesis for same set of parameters. However, the presence of FOPT increases the detection prospects in terms of future observations of stochastic GW.

However, as we increase the scale of FOPT, we get deviations from $T_{\rm RH}=T_n$ resulting in dilution of lepton asymmetry compared to the standard one. To illustrate this, we show three such benchmark points for high scale leptogenesis in table \ref{tab4} and \ref{tab:asym2} which are also consistent with a strong FOPT criteria. As can be seen from table \ref{tab:asym2}, for such high scale FOPT, we have $T_{\rm inf}>T_n$ resulting in $T_{\rm RH} > T_n$. This leads to the dilution of lepton asymmetry by a factor $(T_n/T_{\rm RH})^3$ which is as large as $\sim 10^3$ for the last benchmark point in table \ref{tab:asym2}. Accordingly, the required CP asymmetry parameter needs to be enhanced for such scenarios. Even though we are in weak washout regime, FOPT can also lead to sizeable washout at the end of the phase transition due to the large latent heat released. Such washout effects can be significant if $T_{\rm RH} \gg M_N$. In the benchmark points we have considered, $T_{\rm RH} < M_N$ and hence such washout effects are expected to be smaller. Therefore, in the weak washout regime, high scale FOPT scenario can give successful leptogenesis for the parameter space leading to overproduction of asymmetry in standard leptogenesis.

On the other hand, if we are in the strong washout regime of standard leptogenesis, the FOPT triggered leptogenesis can, in principle, enhance the production of asymmetry if the dilution effects are under control and $T_{\rm RH} \ll M_N$ in order keep the washout processes like inverse decay suppressed. A detailed investigation of this regime along with implications for dark matter will require the relevant Boltzmann equations to be solved numerically, which we leave for future works.
\begin{table}
    \centering
    \begin{tabular}{|c|c|c|c|c|c|c|}
    \hline
     $v_c$  & $T_c$  & v & $v_c/T_c$ & $\lambda_7 (0)$ & $Y'_2 (0)$ &  $\lambda_s (0)$  \\
         (GeV) & (GeV) & (GeV) & &  & $\approx Y'_3 (0)$ & \\
        \hline
        $9.68\times10^7$ & $3.14\times10^7$ & $9.93\times10^7$ & 3.07 & 2.2 & 0.6 &  0.02  \\
       \hline
     $9.52\times10^8$ & $2.05\times10^8$ & $9.76\times10^8$ & 4.63 & 1.2 & 0.6 & 0.02 \\
       \hline
    $9.61\times10^9$ & $2.62\times10^9$ & $9.89\times10^9$ & 3.66 & 1.7 & 0.6 & 0.02 \\
       \hline
           \end{tabular}
    \caption{Benchmark parameters and other details for high scale leptogenesis scenario.}
    \label{tab4}
\end{table}
\begin{table}[]
    \centering
    \begin{tabular}{|c|c|c|c|c|c|}
       \hline $\epsilon_N$  & $T_{\rm RH}$ (GeV) & $T_n$ (GeV)&  $M_{N_{2}} \approx M_{N_3}$ (GeV) & $y_D$ & $\Delta V_{\rm tot}$ (GeV)$^4$\\
        \hline 
        $8.93\times 10^{-7}$  & $9.87 \times 10^6$  & $4.09 \times 10^6$ &  $5.96 \times 10^7$ & $5.62 \times 10^{-6}$ & $3.62\times10^{29}$ \\
        $1.81\times 10^{-7}$  & $7.04 \times 10^7$  & $4.97 \times 10^7$ &  $5.85 \times 10^8$ & $1.76 \times 10^{-5}$ & $9.96\times10^{32}$\\
        $4.40 \times 10^{-5}$ & $8.59 \times 10^8$  & $9.71 \times 10^7$ &  $5.93 \times 10^9$  & $5.60 \times 10^{-5}$ & $2.07\times10^{37}$\\
            \hline
    \end{tabular}
    \caption{CP asymmetry and other relevant details involved in high scale leptogenesis calculation for the hybrid model.}
    \label{tab:asym2}
\end{table}

\begin{figure}[h]
    \centering
    \includegraphics[height=6cm, width = 8cm]{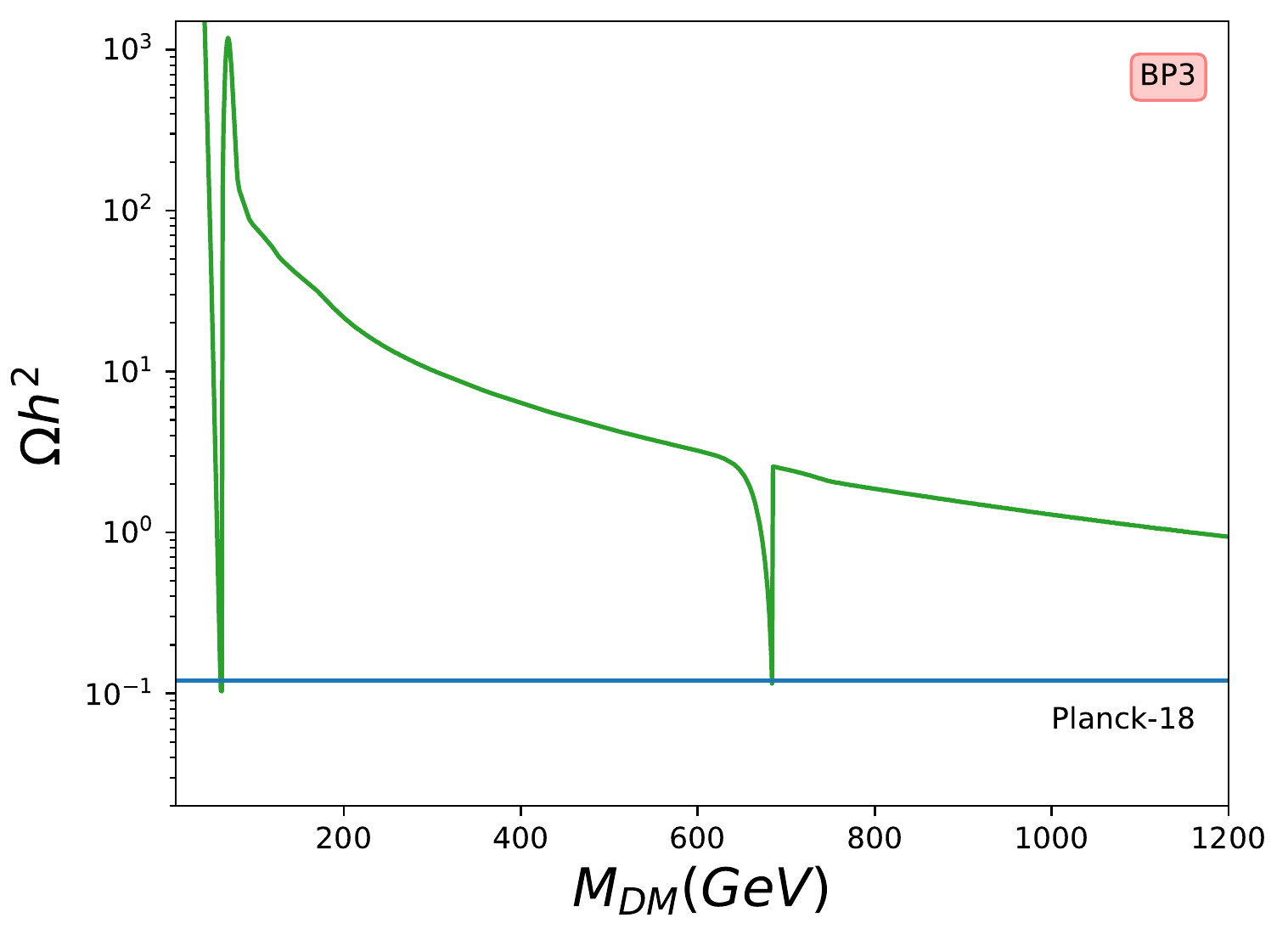}
        \includegraphics[height=6cm, width = 8cm]{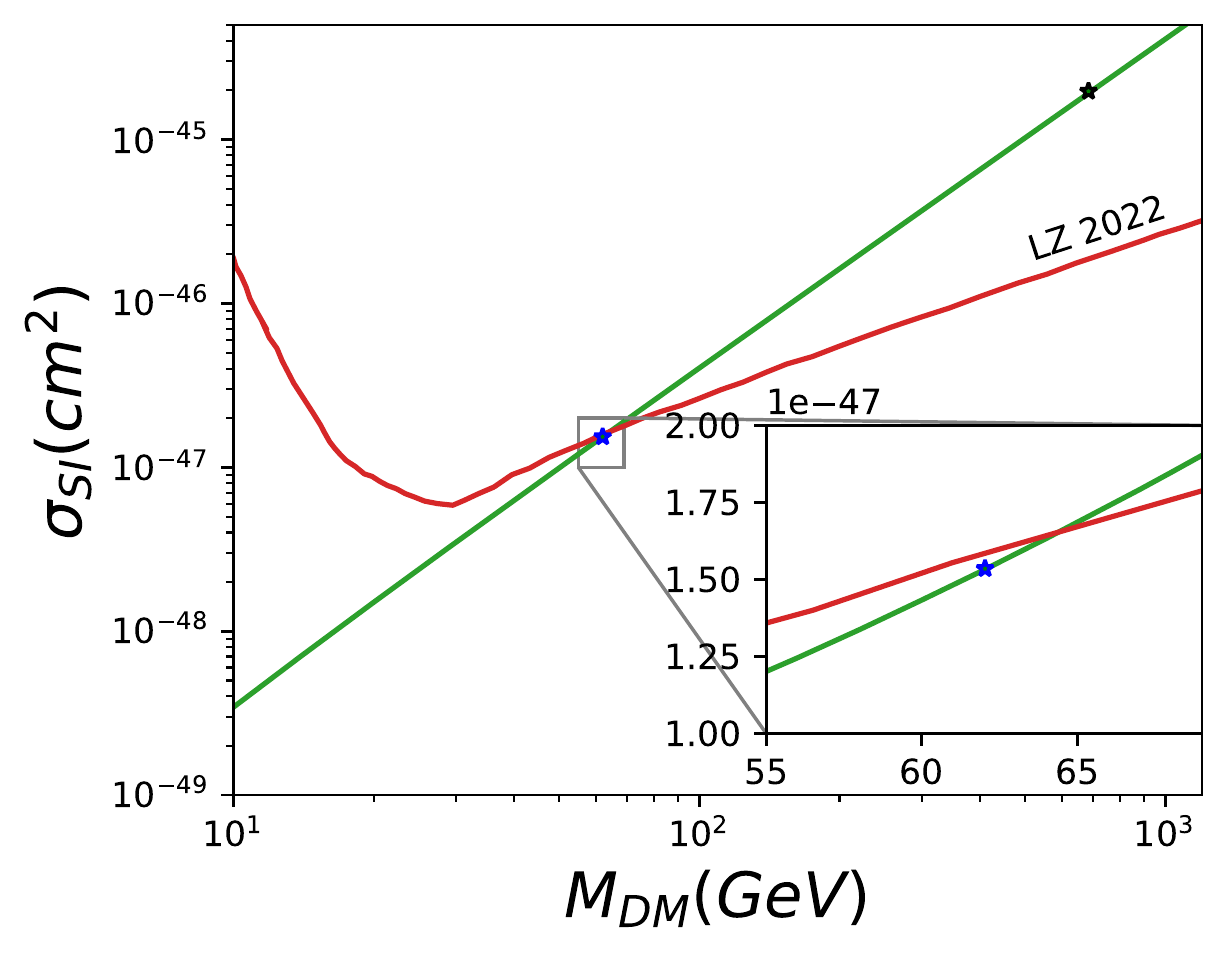}
    \caption{Left panel: WIMP DM relic as a function of its mass for BP3, while considering singlet-SM Higgs mixing to be $\sin{\theta_{hS}} = 0.23$. Right panel: Spin-independent DM-nucleon cross-section as a function of DM mass for the same benchmark choices of parameters considered in left panel.}
    \label{fig:wimp}
\end{figure}

As mentioned earlier, the $Z_2$ odd RHN namely, $N_1$ is the DM candidate which can have light masses due to its small Yukawa couplings with the singlet scalar. It is possible, in principle, to realise either thermal WIMP or non-thermal DM scenario with the latter being popularly known as feebly interacting massive particle (FIMP) \cite{Hall:2009bx}. As $M_{\rm DM} < T_n$, DM can be in equilibrium inside the bubble and undergo thermal freeze-out at a temperature $T_f \sim M_{\rm DM}/20$, if its coupling to the SM bath is sizeable enough. DM can annihilate into SM particles via two possible processes: Yukawa interactions with SM leptons via inert scalar doublet $\eta$ and singlet scalar mediated annihilations into SM particles via singlet-Higgs mixing. Since the scalar doublet $\eta$ is much heavier than the RHNs, the corresponding DM annihilation cross-section remains suppressed compared to the singlet scalar mediated one. Since the singlet scalar mass is small, we can get the desired relic of DM by appropriate tuning of singlet-Higgs mixing. In order to calculate the thermally averaged annihilation cross-sections and solve the Boltzmann equation for DM numerically, we use \texttt{micrOMEGAs} \cite{Belanger:2014vza}. We consider one particular benchmark point namely, BP3 such that the singlet scalar mass is fixed. The relic abundance as a function of DM mass is shown on the left panel of Fig. \ref{fig:wimp}. Even for a considerably large mixing between singlet scalar and the SM Higgs $\sin{\theta_{hS}} = 0.23$, the relic can barely be satisfied at the resonances. Similarly, the stringent direct detection bounds \cite{LUX-ZEPLIN:2022qhg} barely allows the relic satisfying point at the SM Higgs resonance while ruling out the heavier DM mass at singlet scalar resonance. This is due to the fact that, DM Yukawa coupling with singlet scalar is $\langle S \rangle/M_{\rm DM} = v/m_{N1}$ which is very small for this mass range. It should also be noted that the actual singlet-SM Higgs mixing will be $\sin{\theta_{hS}} \sim \lambda_6 v_{\rm ew}/v \ll 0.23$ thereby ruling out the WIMP possibility in this minimal setup.

\begin{figure}[h]
    \centering
    \includegraphics[scale=0.75]{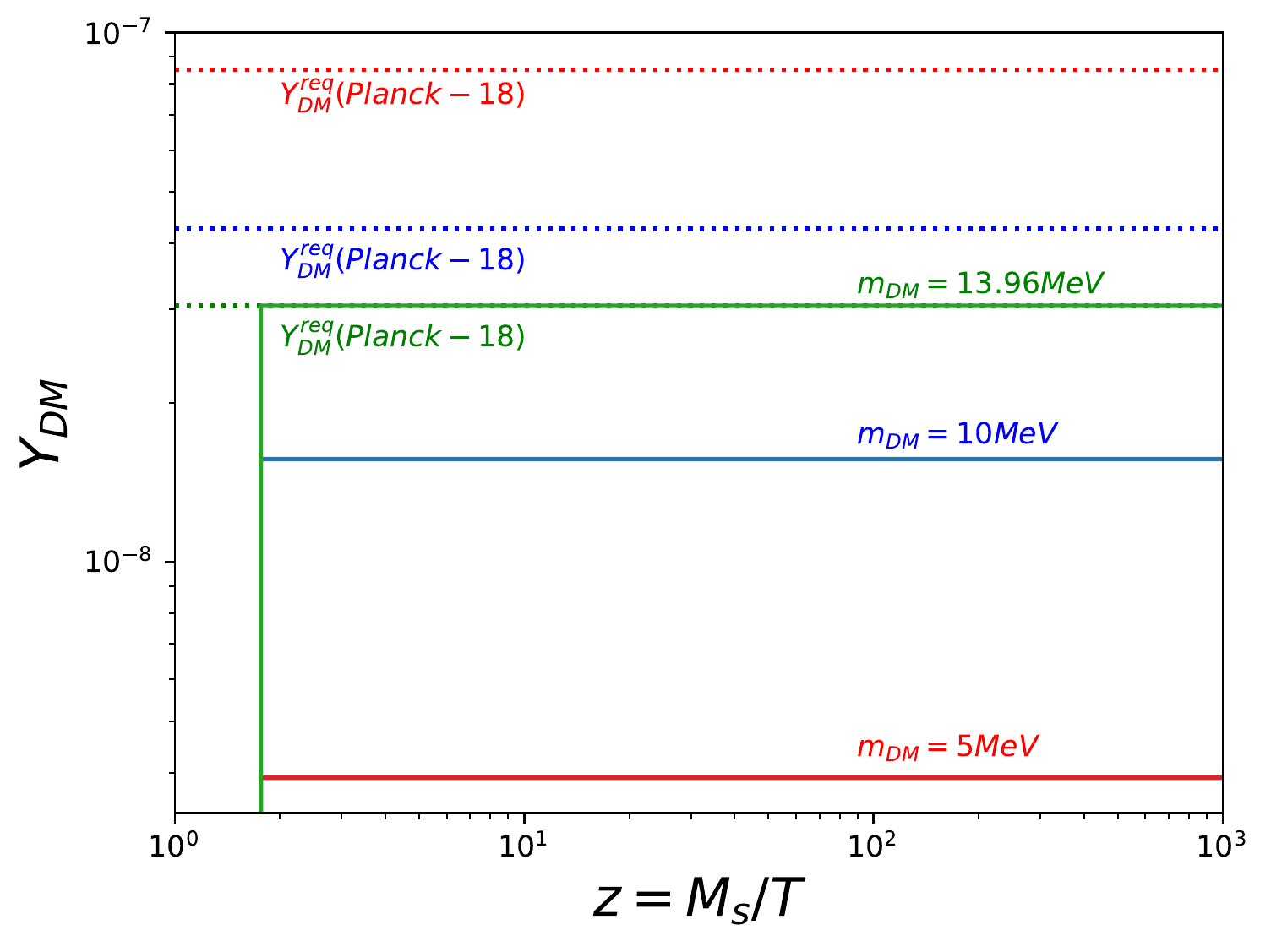}
    \caption{Comoving density of FIMP DM for BP3}
    \label{fig:fimp}
\end{figure}

We finally consider the FIMP DM possibility by considering singlet scalar decay after the phase transition. The corresponding Boltzmann equations can be written as
\begin{equation}
    \frac{dY_{DM}}{dz}=\frac{2}{z \bf{H}}  \Gamma_{sN1}  Y_s, \,\,
    \frac{dY_s}{dz}=-\frac{1}{z \bf{H}} ( \Gamma_{sN1} + \Gamma_{sh} ) Y_s,
\end{equation}
with $z=M_s/T$ and assuming the relativistic dof to be constant, which is valid at temperatures above electroweak scale. While DM is produced only from the decay of singlet scalar, the latter can decay into SM Higgs as well. The corresponding decay widths are
\begin{equation}
    \Gamma_{sN1}=\frac{1}{16 \pi}y_1^2 M_s \left (1-\frac{4M_{DM}^2}{M_s^2}\right)^{3/2}, \,\,
    \Gamma_{sh} =\frac{y_{shh}^2}{16 \pi M_s}\left (1-\frac{4M_{h}^2}{M_s^2}\right )^{1/2}, 
\end{equation}
where $y_{shh} \sim \lambda_6 v$. Similar to the case of heavy RHNs $N_{2,3}$, the singlet scalar also acquires a large abundance inside the bubble at nucleation temperature, close to the equilibrium comoving number density without Boltzmann suppression. Considering this to be the initial comoving abundance of S and taking appropriate partial decay width of S into SM Higgs, we solve the above Boltzmann equations for different DM masses and find that for DM mass around 14 MeV, the correct relic is satisfied if other parameters are fixed as in BP3 discussed before. Since singlet decay width into SM Higgs is substantial, we require somewhat large FIMP mass to satisfy the correct DM relic. This also leads to instantaneous freeze-in shortly after nucleation temperature $T \sim T_*$ as larger mass corresponds to larger Yukawa coupling of DM with singlet.

\section{Conclusion}
\label{sec6}
We have studied the possibility of getting dark matter and low scale leptogenesis from a supercooled first order phase transition driven by a singlet scalar around TeV scale. The right handed neutrinos responsible for generating lepton asymmetry via decay and dark matter acquire masses by crossing the relativistic bubble walls which arise as a result of the FOPT. This also leads to a large abundance of RHN in true vacuum inside the bubble sufficient for generating the required lepton asymmetry without washout or Boltzmann suppression. The dark matter is lighter than the nucleation temperature and hence can remain in equilibrium inside the bubble with its relic determined by thermal freeze-out at later stages. In order to implement the idea, we first consider a conformal version of the scotogenic model such that along with dark matter and right handed neutrino generating radiative light neutrino masses, we also have a strong supercooling to bring the resulting gravitational wave amplitude within near future experiment's sensitivity. While a strong supercooled FOPT is possible, the hierarchy of the additional field content of the model does not allow the realisation of leptogenesis from RHN decay. We then consider a hybrid scenario with the same field content but different seesaw realisation to show correct DM phenomenology from the lightest RHN while the heavier two RHNs can lead to successful TeV scale resonant leptogenesis. The light neutrino mass arises from a hybrid seesaw mechanism involving both type I and radiative origin. As the FOPT details remain more or less similar to the conformal scotogenic model, we can probe this hybrid model in near future GW experiments like the LISA experiment. Due to TeV scale RHN and additional scalars, the model can also have complementary detection prospects at intensity and energy frontier experiments.

\appendix

\section{Renormalisation Group Evolution Equations}
\label{appen1}
The relevant RGE equations for the model parameters are \cite{Bhattacharya:2019tqq}

    \begin{equation*}
        \frac{d \lambda_s}{dt}=\frac{1}{16\pi^2}(20\lambda_s^2+ 2\lambda_6^2+2\lambda_7^2+8\lambda_s {\rm Tr}[Y'^{\dagger} Y']-{\rm Tr}[Y'^{\dagger} Y'Y'^{\dagger} Y'])
    \end{equation*}
    \begin{equation*}
    \frac{d \lambda_2}{dt}=\frac{1}{16\pi^2}(12\lambda_2^2+2\lambda_7^2+3g_1^2/4 +9g_2^2/4 +3g_1^2g_2^2/2)
    \end{equation*}
   \begin{equation*}
   \frac{d \lambda_7}{dt}=\frac{1}{16\pi^2}(4\lambda_7^2+6\lambda_2 \lambda_7+8\lambda_s\lambda_7 +4\lambda_7 {\rm Tr}[Y'^{\dagger} Y'])
   \end{equation*}
   \begin{equation*}
   \frac{d \lambda_6}{dt}=\frac{1}{16\pi^2}(4\lambda_6^2+6\lambda_6 y_t^2+8\lambda_s\lambda_6 +4\lambda_6 {\rm Tr}[Y'^{\dagger} Y'])
   \end{equation*}
   \begin{equation*}
   \frac{d Y'}{dt}=\frac{1}{16\pi^2}(4Y'^3+2Y' {\rm Tr}[Y'^{\dagger} Y'])
   \end{equation*}
   \begin{equation*}
   \frac{d g_1}{dt}=\frac{1}{16\pi^2}(7g_1^3)
   \end{equation*}
   \begin{equation*}
   \frac{d g_2}{dt}=\frac{1}{16\pi^2}(-3g_2^3)
   \end{equation*}
   \begin{equation*}
   \frac{d y_t}{dt}=\frac{1}{16\pi^2}(9y_t^3/2-y_t(17g_1^2/12+9g_2^2/4))
   \end{equation*}
   
\section{Fitting of the finite temperature potential}
\label{appen2}
\begin{figure}[h]
    \centering
    \includegraphics[scale=0.5]{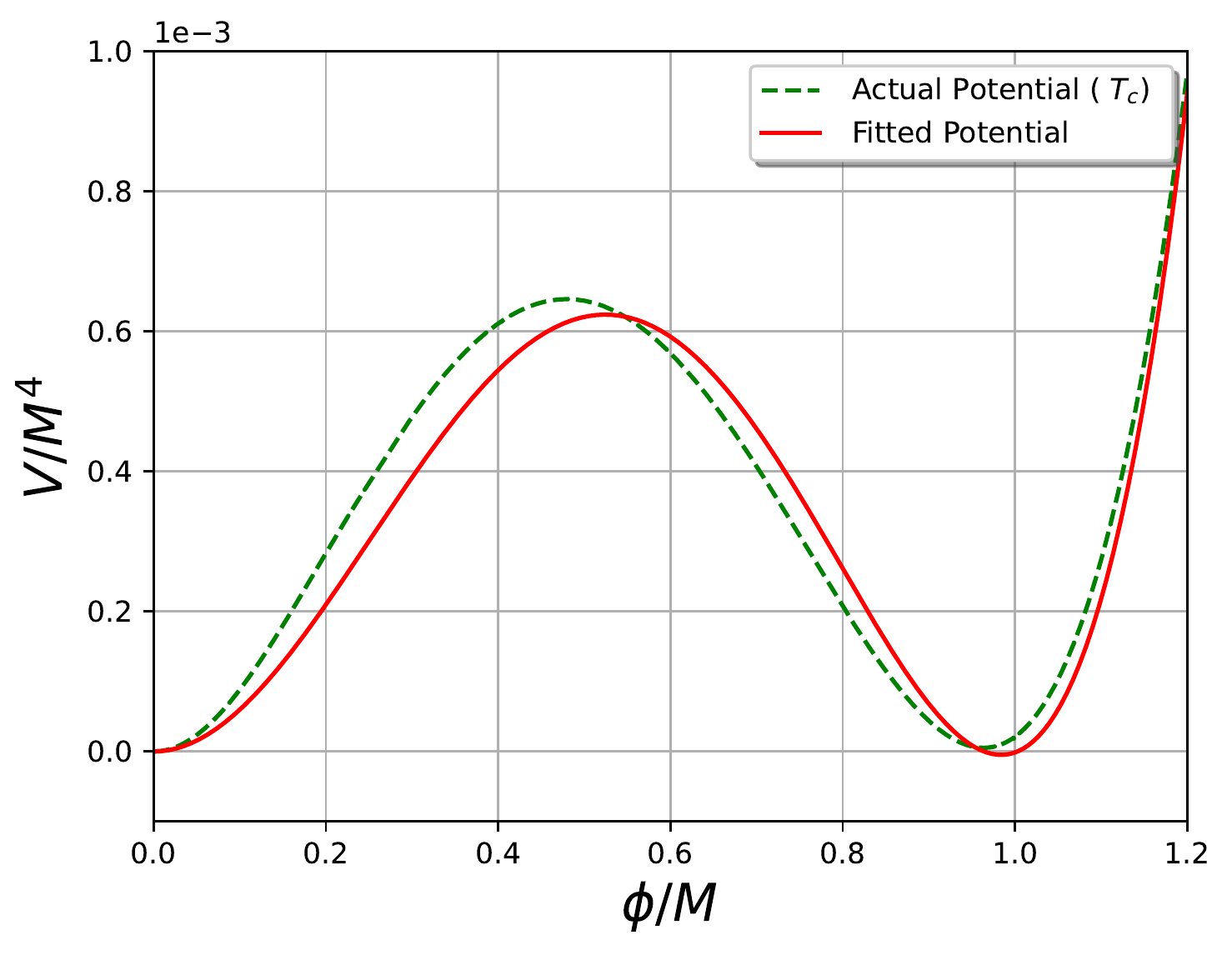}
        \includegraphics[scale=0.5]{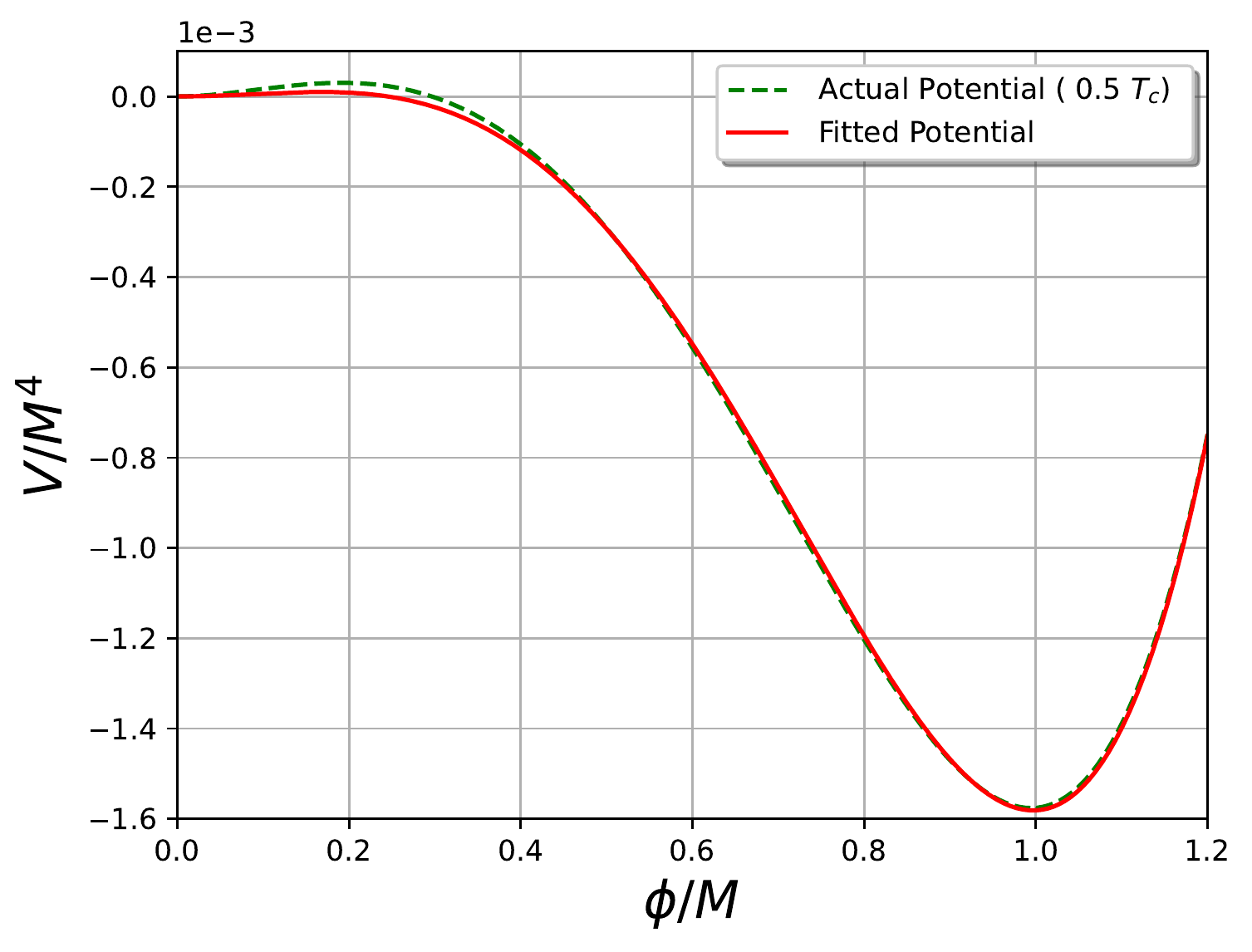} \\
            \includegraphics[scale=0.5]{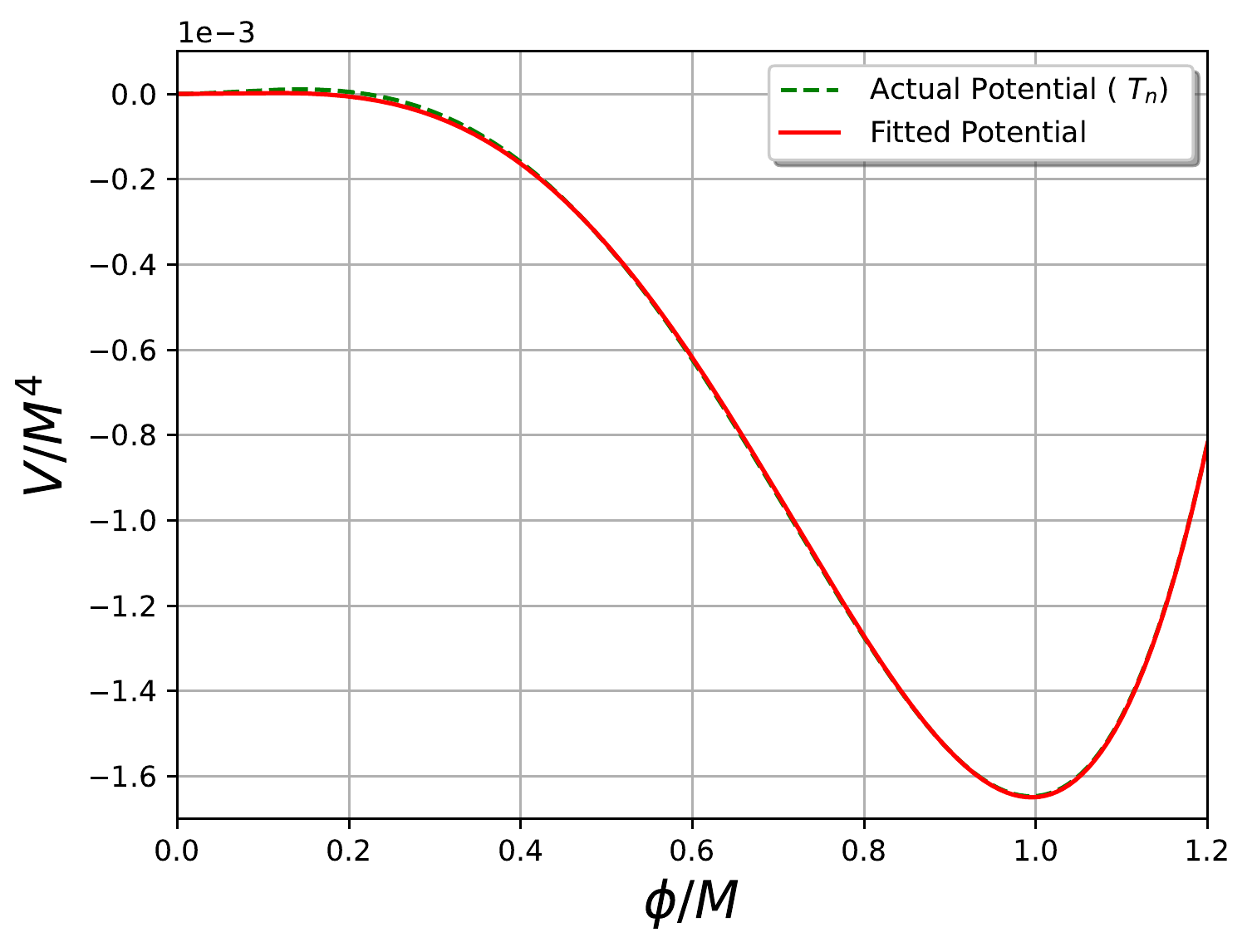}
                \includegraphics[scale=0.5]{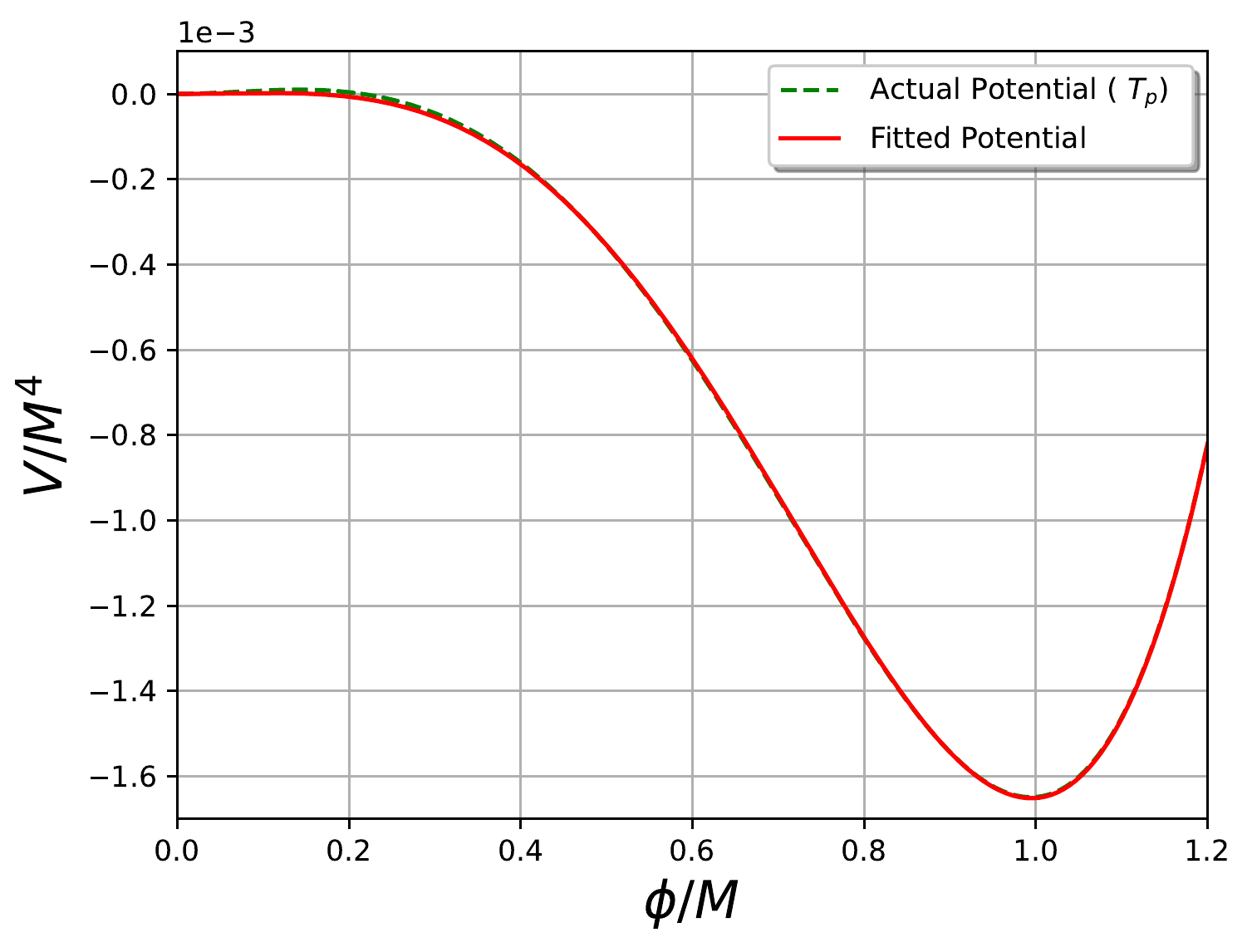}
        \caption{Comparison between the actual thermal potential and the fit at different temperatures.}        
    \label{fig:fit}
\end{figure}

The  generic form of quartic and logarithmic potential can be written as\cite{Adams:1993zs}
\begin{equation}
    V(\phi)=(2A-B)\sigma^2 \phi^2 -A \phi^4 + B \phi^4 \ln{\frac{\phi^2}{\sigma^2}}
\end{equation}
The above expression as the effective potential can be used to calculate the semi-analytical expression of the Euclidean action in terms of the parameters of the potential. We can use the effective potential with consideration of the running coupling effect, one-loop thermal contribution, and Daisy corrections, which can then be fitted well to the generic potential as shown in Fig. \ref{fig:fit} for BP1.
 \\
The bounce equation of motion for the above generic potential in three dimensions is
\begin{equation}
    \frac{d^2\phi}{dr^2}+\frac{2}{r}\frac{d\phi}{dr}=\frac{d V}{d\phi}
\end{equation}
We obtain the derivative of the potential as
\begin{equation}
    \frac{d V}{d \phi}=4B\sigma^3 \Bigl\{\frac{2A-B}{2B}\Bigl(\frac{\phi}{\sigma}-\frac{\phi^3}{\sigma^3}\Bigl)+\frac{\phi^3}{\sigma^3} \ln{\frac{\phi^2}{\sigma^2}}\Bigl\}
\end{equation}
Now using a scaling transformation ($\Phi=\phi/\sigma, \xi=2\sigma \sqrt{B}r$), we reduce the three parameters differential equation (DE) to one parameter DE given by
\begin{equation}
    \frac{d^2\Phi}{d\xi^2}+\frac{2}{\xi}\frac{d\Phi}{d\xi}=\delta(\Phi-\Phi^3)+\Phi^3\ln{\Phi^2}
\end{equation}
where $\delta=\frac{2A-B}{2B}$. Following the approach of \cite{Adams:1993zs}, the action in three dimensions, calculated in a semi-analytical manner, is found to be
\begin{equation}
    S_3=\frac{16 \pi \sigma I^3}{3(1-2\delta)^2}\Bigl(\frac{2}{B}\Bigl)^{1/2}(2\delta)^{n_\mu}\Bigl\{1+\mu_1\delta+\mu_2\delta^2+\mu_3\delta^3\Bigl\}
\end{equation}
where, I=0.4199, $n_\mu=0.557$, $\mu_1=4.2719$, $\mu_2=-14.5908$ and $\mu_3=12.0940$. We have used it in our numerical analysis.


\begin{thebibliography}{100}

\bibitem{Aghanim:2018eyx}
{\scshape Planck} collaboration, \emph{{Planck 2018 results. VI. Cosmological
  parameters}},  \href{https://arxiv.org/abs/1807.06209}{{\ttfamily
  1807.06209}}.

\bibitem{Zyla:2020zbs}
{\scshape Particle Data Group} collaboration, \emph{{Review of Particle
  Physics}}, \href{https://doi.org/10.1093/ptep/ptaa104}{\emph{PTEP} {\bfseries
  2020} (2020) 083C01}.

\bibitem{Sakharov:1967dj}
A.~D. Sakharov, \emph{{Violation of CP Invariance, C asymmetry, and baryon
  asymmetry of the universe}},
  \href{https://doi.org/10.1070/PU1991v034n05ABEH002497}{\emph{Pisma Zh. Eksp.
  Teor. Fiz.} {\bfseries 5} (1967) 32}.

\bibitem{Weinberg:1979bt}
S.~Weinberg, \emph{{Cosmological Production of Baryons}},
  \href{https://doi.org/10.1103/PhysRevLett.42.850}{\emph{Phys. Rev. Lett.}
  {\bfseries 42} (1979) 850}.

\bibitem{Kolb:1979qa}
E.~W. Kolb and S.~Wolfram, \emph{{Baryon Number Generation in the Early
  Universe}}, \href{https://doi.org/10.1016/0550-3213(80)90167-4,
  10.1016/0550-3213(82)90012-8}{\emph{Nucl. Phys.} {\bfseries B172} (1980)
  224}.

\bibitem{Fukugita:1986hr}
M.~Fukugita and T.~Yanagida, \emph{{Baryogenesis Without Grand Unification}},
  \href{https://doi.org/10.1016/0370-2693(86)91126-3}{\emph{Phys. Lett.}
  {\bfseries B174} (1986) 45}.

\bibitem{Kuzmin:1985mm}
V.~A. Kuzmin, V.~A. Rubakov and M.~E. Shaposhnikov, \emph{{On the Anomalous
  Electroweak Baryon Number Nonconservation in the Early Universe}},
  \href{https://doi.org/10.1016/0370-2693(85)91028-7}{\emph{Phys. Lett.}
  {\bfseries 155B} (1985) 36}.

\bibitem{Davidson:2002qv}
S.~Davidson and A.~Ibarra, \emph{{A Lower bound on the right-handed neutrino
  mass from leptogenesis}},
  \href{https://doi.org/10.1016/S0370-2693(02)01735-5}{\emph{Phys. Lett.}
  {\bfseries B535} (2002) 25}
  [\href{https://arxiv.org/abs/hep-ph/0202239}{{\ttfamily hep-ph/0202239}}].

\bibitem{Akhmedov:1998qx}
E.~K. Akhmedov, V.~A. Rubakov and A.~Y. Smirnov, \emph{{Baryogenesis via
  neutrino oscillations}},
  \href{https://doi.org/10.1103/PhysRevLett.81.1359}{\emph{Phys. Rev. Lett.}
  {\bfseries 81} (1998) 1359}
  [\href{https://arxiv.org/abs/hep-ph/9803255}{{\ttfamily hep-ph/9803255}}].

\bibitem{Asaka:2005pn}
T.~Asaka and M.~Shaposhnikov, \emph{{The $\nu$MSM, dark matter and baryon
  asymmetry of the universe}},
  \href{https://doi.org/10.1016/j.physletb.2005.06.020}{\emph{Phys. Lett. B}
  {\bfseries 620} (2005) 17}
  [\href{https://arxiv.org/abs/hep-ph/0505013}{{\ttfamily hep-ph/0505013}}].

\bibitem{Abada:2018oly}
A.~Abada, G.~Arcadi, V.~Domcke, M.~Drewes, J.~Klaric and M.~Lucente,
  \emph{{Low-scale leptogenesis with three heavy neutrinos}},
  \href{https://doi.org/10.1007/JHEP01(2019)164}{\emph{JHEP} {\bfseries 01}
  (2019) 164} [\href{https://arxiv.org/abs/1810.12463}{{\ttfamily
  1810.12463}}].

\bibitem{Drewes:2021nqr}
M.~Drewes, Y.~Georis and J.~Klari\'c, \emph{{Mapping the Viable Parameter Space
  for Testable Leptogenesis}},
  \href{https://doi.org/10.1103/PhysRevLett.128.051801}{\emph{Phys. Rev. Lett.}
  {\bfseries 128} (2022) 051801}
  [\href{https://arxiv.org/abs/2106.16226}{{\ttfamily 2106.16226}}].

\bibitem{LeDall:2014too}
M.~Le~Dall and A.~Ritz, \emph{{Leptogenesis and the Higgs Portal}},
  \href{https://doi.org/10.1103/PhysRevD.90.096002}{\emph{Phys. Rev. D}
  {\bfseries 90} (2014) 096002}
  [\href{https://arxiv.org/abs/1408.2498}{{\ttfamily 1408.2498}}].

\bibitem{Alanne:2018brf}
T.~Alanne, T.~Hugle, M.~Platscher and K.~Schmitz, \emph{{Low-scale leptogenesis
  assisted by a real scalar singlet}},
  \href{https://doi.org/10.1088/1475-7516/2019/03/037}{\emph{JCAP} {\bfseries
  03} (2019) 037} [\href{https://arxiv.org/abs/1812.04421}{{\ttfamily
  1812.04421}}].

\bibitem{Hambye:2009pw}
T.~Hambye, F.~S. Ling, L.~Lopez~Honorez and J.~Rocher, \emph{{Scalar Multiplet
  Dark Matter}}, \href{https://doi.org/10.1007/JHEP05(2010)066,
  10.1088/1126-6708/2009/07/090}{\emph{JHEP} {\bfseries 07} (2009) 090}
  [\href{https://arxiv.org/abs/0903.4010}{{\ttfamily 0903.4010}}].

\bibitem{Racker:2013lua}
J.~Racker, \emph{{Mass bounds for baryogenesis from particle decays and the
  inert doublet model}},
  \href{https://doi.org/10.1088/1475-7516/2014/03/025}{\emph{JCAP} {\bfseries
  03} (2014) 025} [\href{https://arxiv.org/abs/1308.1840}{{\ttfamily
  1308.1840}}].

\bibitem{Clarke:2015hta}
J.~D. Clarke, R.~Foot and R.~R. Volkas, \emph{{Natural leptogenesis and
  neutrino masses with two Higgs doublets}},
  \href{https://doi.org/10.1103/PhysRevD.92.033006}{\emph{Phys. Rev. D}
  {\bfseries 92} (2015) 033006}
  [\href{https://arxiv.org/abs/1505.05744}{{\ttfamily 1505.05744}}].

\bibitem{Hugle:2018qbw}
T.~Hugle, M.~Platscher and K.~Schmitz, \emph{{Low-Scale Leptogenesis in the
  Scotogenic Neutrino Mass Model}},
  \href{https://doi.org/10.1103/PhysRevD.98.023020}{\emph{Phys. Rev. D}
  {\bfseries 98} (2018) 023020}
  [\href{https://arxiv.org/abs/1804.09660}{{\ttfamily 1804.09660}}].

\bibitem{Borah:2018rca}
D.~Borah, P.~S.~B. Dev and A.~Kumar, \emph{{TeV scale leptogenesis, inflaton
  dark matter and neutrino mass in a scotogenic model}},
  \href{https://doi.org/10.1103/PhysRevD.99.055012}{\emph{Phys. Rev. D}
  {\bfseries 99} (2019) 055012}
  [\href{https://arxiv.org/abs/1810.03645}{{\ttfamily 1810.03645}}].

\bibitem{Mahanta:2019gfe}
D.~Mahanta and D.~Borah, \emph{{Fermion dark matter with $N_2$ leptogenesis in
  minimal scotogenic model}},
  \href{https://doi.org/10.1088/1475-7516/2019/11/021}{\emph{JCAP} {\bfseries
  11} (2019) 021} [\href{https://arxiv.org/abs/1906.03577}{{\ttfamily
  1906.03577}}].

\bibitem{Mahanta:2019sfo}
D.~Mahanta and D.~Borah, \emph{{TeV Scale Leptogenesis with Dark Matter in
  Non-standard Cosmology}},
  \href{https://doi.org/10.1088/1475-7516/2020/04/032}{\emph{JCAP} {\bfseries
  04} (2020) 032} [\href{https://arxiv.org/abs/1912.09726}{{\ttfamily
  1912.09726}}].

\bibitem{Sarma:2020msa}
L.~Sarma, P.~Das and M.~K. Das, \emph{{Scalar dark matter and leptogenesis in
  the minimal scotogenic model}},
  \href{https://doi.org/10.1016/j.nuclphysb.2020.115300}{\emph{Nucl. Phys. B}
  {\bfseries 963} (2021) 115300}
  [\href{https://arxiv.org/abs/2004.13762}{{\ttfamily 2004.13762}}].

\bibitem{Borah:2020ivi}
D.~Borah, A.~Dasgupta and D.~Mahanta, \emph{{Dark sector assisted low scale
  leptogenesis from three body decay}},
  \href{https://doi.org/10.1103/PhysRevD.105.015015}{\emph{Phys. Rev. D}
  {\bfseries 105} (2022) 015015}
  [\href{https://arxiv.org/abs/2008.10627}{{\ttfamily 2008.10627}}].

\bibitem{Pilaftsis:1998pd}
A.~Pilaftsis, \emph{{Heavy Majorana neutrinos and baryogenesis}},
  \href{https://doi.org/10.1142/S0217751X99000932}{\emph{Int. J. Mod. Phys. A}
  {\bfseries 14} (1999) 1811}
  [\href{https://arxiv.org/abs/hep-ph/9812256}{{\ttfamily hep-ph/9812256}}].

\bibitem{Pilaftsis:2003gt}
A.~Pilaftsis and T.~E.~J. Underwood, \emph{{Resonant leptogenesis}},
  \href{https://doi.org/10.1016/j.nuclphysb.2004.05.029}{\emph{Nucl. Phys.}
  {\bfseries B692} (2004) 303}
  [\href{https://arxiv.org/abs/hep-ph/0309342}{{\ttfamily hep-ph/0309342}}].

\bibitem{Chun:2017spz}
E.~J. Chun et~al., \emph{{Probing Leptogenesis}},
  \href{https://doi.org/10.1142/S0217751X18420058}{\emph{Int. J. Mod. Phys. A}
  {\bfseries 33} (2018) 1842005}
  [\href{https://arxiv.org/abs/1711.02865}{{\ttfamily 1711.02865}}].

\bibitem{Dror:2019syi}
J.~A. Dror, T.~Hiramatsu, K.~Kohri, H.~Murayama and G.~White, \emph{{Testing
  the Seesaw Mechanism and Leptogenesis with Gravitational Waves}},
  \href{https://doi.org/10.1103/PhysRevLett.124.041804}{\emph{Phys. Rev. Lett.}
  {\bfseries 124} (2020) 041804}
  [\href{https://arxiv.org/abs/1908.03227}{{\ttfamily 1908.03227}}].

\bibitem{Blasi:2020wpy}
S.~Blasi, V.~Brdar and K.~Schmitz, \emph{{Fingerprint of low-scale leptogenesis
  in the primordial gravitational-wave spectrum}},
  \href{https://doi.org/10.1103/PhysRevResearch.2.043321}{\emph{Phys. Rev.
  Res.} {\bfseries 2} (2020) 043321}
  [\href{https://arxiv.org/abs/2004.02889}{{\ttfamily 2004.02889}}].

\bibitem{Fornal:2020esl}
B.~Fornal and B.~Shams Es~Haghi, \emph{{Baryon and Lepton Number Violation from
  Gravitational Waves}},
  \href{https://doi.org/10.1103/PhysRevD.102.115037}{\emph{Phys. Rev. D}
  {\bfseries 102} (2020) 115037}
  [\href{https://arxiv.org/abs/2008.05111}{{\ttfamily 2008.05111}}].

\bibitem{Samanta:2020cdk}
R.~Samanta and S.~Datta, \emph{{Gravitational wave complementarity and impact
  of NANOGrav data on gravitational leptogenesis: cosmic strings}},
  \href{https://arxiv.org/abs/2009.13452}{{\ttfamily 2009.13452}}.

\bibitem{Barman:2022yos}
B.~Barman, D.~Borah, A.~Dasgupta and A.~Ghoshal, \emph{{Probing high scale
  Dirac leptogenesis via gravitational waves from domain walls}},
  \href{https://doi.org/10.1103/PhysRevD.106.015007}{\emph{Phys. Rev. D}
  {\bfseries 106} (2022) 015007}
  [\href{https://arxiv.org/abs/2205.03422}{{\ttfamily 2205.03422}}].

\bibitem{Baldes:2021vyz}
I.~Baldes, S.~Blasi, A.~Mariotti, A.~Sevrin and K.~Turbang, \emph{{Baryogenesis
  via relativistic bubble expansion}},
  \href{https://doi.org/10.1103/PhysRevD.104.115029}{\emph{Phys. Rev. D}
  {\bfseries 104} (2021) 115029}
  [\href{https://arxiv.org/abs/2106.15602}{{\ttfamily 2106.15602}}].

\bibitem{Azatov:2021irb}
A.~Azatov, M.~Vanvlasselaer and W.~Yin, \emph{{Baryogenesis via relativistic
  bubble walls}}, \href{https://doi.org/10.1007/JHEP10(2021)043}{\emph{JHEP}
  {\bfseries 10} (2021) 043}
  [\href{https://arxiv.org/abs/2106.14913}{{\ttfamily 2106.14913}}].

\bibitem{Huang:2022vkf}
P.~Huang and K.-P. Xie, \emph{{Leptogenesis triggered by a first-order phase
  transition}},  \href{https://arxiv.org/abs/2206.04691}{{\ttfamily
  2206.04691}}.

\bibitem{Dasgupta:2022isg}
A.~Dasgupta, P.~S.~B. Dev, A.~Ghoshal and A.~Mazumdar, \emph{{Gravitational
  Wave Pathway to Testable Leptogenesis}},
  \href{https://arxiv.org/abs/2206.07032}{{\ttfamily 2206.07032}}.

\bibitem{Yuan:2021ebu}
C.~Yuan, R.~Brito and V.~Cardoso, \emph{{Probing ultralight dark matter with
  future ground-based gravitational-wave detectors}},
  \href{https://doi.org/10.1103/PhysRevD.104.044011}{\emph{Phys. Rev. D}
  {\bfseries 104} (2021) 044011}
  [\href{https://arxiv.org/abs/2106.00021}{{\ttfamily 2106.00021}}].

\bibitem{Tsukada:2020lgt}
L.~Tsukada, R.~Brito, W.~E. East and N.~Siemonsen, \emph{{Modeling and
  searching for a stochastic gravitational-wave background from ultralight
  vector bosons}},
  \href{https://doi.org/10.1103/PhysRevD.103.083005}{\emph{Phys. Rev. D}
  {\bfseries 103} (2021) 083005}
  [\href{https://arxiv.org/abs/2011.06995}{{\ttfamily 2011.06995}}].

\bibitem{Chatrchyan:2020pzh}
A.~Chatrchyan and J.~Jaeckel, \emph{{Gravitational waves from the fragmentation
  of axion-like particle dark matter}},
  \href{https://doi.org/10.1088/1475-7516/2021/02/003}{\emph{JCAP} {\bfseries
  02} (2021) 003} [\href{https://arxiv.org/abs/2004.07844}{{\ttfamily
  2004.07844}}].

\bibitem{Bian:2021vmi}
L.~Bian, X.~Liu and K.-P. Xie, \emph{{Probing superheavy dark matter with
  gravitational waves}},
  \href{https://doi.org/10.1007/JHEP11(2021)175}{\emph{JHEP} {\bfseries 11}
  (2021) 175} [\href{https://arxiv.org/abs/2107.13112}{{\ttfamily
  2107.13112}}].

\bibitem{Samanta:2021mdm}
R.~Samanta and F.~R. Urban, \emph{{Testing Super Heavy Dark Matter from
  Primordial Black Holes with Gravitational Waves}},
  \href{https://arxiv.org/abs/2112.04836}{{\ttfamily 2112.04836}}.

\bibitem{Borah:2022byb}
D.~Borah, S.~J. Das, A.~K. Saha and R.~Samanta, \emph{{Probing Miracle-less
  WIMP Dark Matter via Gravitational Waves Spectral Shapes}},
  \href{https://arxiv.org/abs/2202.10474}{{\ttfamily 2202.10474}}.

\bibitem{Azatov:2021ifm}
A.~Azatov, M.~Vanvlasselaer and W.~Yin, \emph{{Dark Matter production from
  relativistic bubble walls}},
  \href{https://doi.org/10.1007/JHEP03(2021)288}{\emph{JHEP} {\bfseries 03}
  (2021) 288} [\href{https://arxiv.org/abs/2101.05721}{{\ttfamily
  2101.05721}}].

\bibitem{Azatov:2022tii}
A.~Azatov, G.~Barni, S.~Chakraborty, M.~Vanvlasselaer and W.~Yin,
  \emph{{Ultra-relativistic bubbles from the simplest Higgs portal and their
  cosmological consequences}},
  \href{https://arxiv.org/abs/2207.02230}{{\ttfamily 2207.02230}}.

\bibitem{Baldes:2022oev}
I.~Baldes, Y.~Gouttenoire and F.~Sala, \emph{{Hot and Heavy Dark Matter from
  Supercooling}},  \href{https://arxiv.org/abs/2207.05096}{{\ttfamily
  2207.05096}}.

\bibitem{LUX-ZEPLIN:2022qhg}
{\scshape LUX-ZEPLIN} collaboration, \emph{{First Dark Matter Search Results
  from the LUX-ZEPLIN (LZ) Experiment}},
  \href{https://arxiv.org/abs/2207.03764}{{\ttfamily 2207.03764}}.

\bibitem{Arakawa:2021wgz}
J.~Arakawa, A.~Rajaraman and T.~M.~P. Tait, \emph{{Annihilogenesis}},
  \href{https://doi.org/10.1007/JHEP08(2022)078}{\emph{JHEP} {\bfseries 08}
  (2022) 078} [\href{https://arxiv.org/abs/2109.13941}{{\ttfamily
  2109.13941}}].

\bibitem{Ahmadvand:2021vxs}
M.~Ahmadvand, \emph{{Filtered asymmetric dark matter during the Peccei-Quinn
  phase transition}},
  \href{https://doi.org/10.1007/JHEP10(2021)109}{\emph{JHEP} {\bfseries 10}
  (2021) 109} [\href{https://arxiv.org/abs/2108.00958}{{\ttfamily
  2108.00958}}].

\bibitem{Ma:2006km}
E.~Ma, \emph{{Verifiable radiative seesaw mechanism of neutrino mass and dark
  matter}}, \href{https://doi.org/10.1103/PhysRevD.73.077301}{\emph{Phys. Rev.
  D} {\bfseries 73} (2006) 077301}
  [\href{https://arxiv.org/abs/hep-ph/0601225}{{\ttfamily hep-ph/0601225}}].

\bibitem{Ahriche:2016cio}
A.~Ahriche, K.~L. McDonald and S.~Nasri, \emph{{The Scale-Invariant Scotogenic
  Model}}, \href{https://doi.org/10.1007/JHEP06(2016)182}{\emph{JHEP}
  {\bfseries 06} (2016) 182}
  [\href{https://arxiv.org/abs/1604.05569}{{\ttfamily 1604.05569}}].

\bibitem{Merle:2015ica}
A.~Merle and M.~Platscher, \emph{{Running of radiative neutrino masses: the
  scotogenic model \textemdash{} revisited}},
  \href{https://doi.org/10.1007/JHEP11(2015)148}{\emph{JHEP} {\bfseries 11}
  (2015) 148} [\href{https://arxiv.org/abs/1507.06314}{{\ttfamily
  1507.06314}}].

\bibitem{Borah:2020wut}
D.~Borah, A.~Dasgupta, K.~Fujikura, S.~K. Kang and D.~Mahanta,
  \emph{{Observable Gravitational Waves in Minimal Scotogenic Model}},
  \href{https://doi.org/10.1088/1475-7516/2020/08/046}{\emph{JCAP} {\bfseries
  08} (2020) 046} [\href{https://arxiv.org/abs/2003.02276}{{\ttfamily
  2003.02276}}].

\bibitem{Dolan:1973qd}
L.~Dolan and R.~Jackiw, \emph{{Symmetry Behavior at Finite Temperature}},
  \href{https://doi.org/10.1103/PhysRevD.9.3320}{\emph{Phys. Rev. D} {\bfseries
  9} (1974) 3320}.

\bibitem{Quiros:1999jp}
M.~Quiros, \emph{{Finite temperature field theory and phase transitions}},  in
  \emph{{ICTP Summer School in High-Energy Physics and Cosmology}},
  pp.~187--259, 1, 1999, \href{https://arxiv.org/abs/hep-ph/9901312}{{\ttfamily
  hep-ph/9901312}}.

\bibitem{Wainwright:2011qy}
C.~Wainwright, S.~Profumo and M.~J. Ramsey-Musolf, \emph{{Gravity Waves from a
  Cosmological Phase Transition: Gauge Artifacts and Daisy Resummations}},
  \href{https://doi.org/10.1103/PhysRevD.84.023521}{\emph{Phys. Rev. D}
  {\bfseries 84} (2011) 023521}
  [\href{https://arxiv.org/abs/1104.5487}{{\ttfamily 1104.5487}}].

\bibitem{Wainwright:2012zn}
C.~L. Wainwright, S.~Profumo and M.~J. Ramsey-Musolf, \emph{{Phase Transitions
  and Gauge Artifacts in an Abelian Higgs Plus Singlet Model}},
  \href{https://doi.org/10.1103/PhysRevD.86.083537}{\emph{Phys. Rev. D}
  {\bfseries 86} (2012) 083537}
  [\href{https://arxiv.org/abs/1204.5464}{{\ttfamily 1204.5464}}].

\bibitem{Coleman:1973jx}
S.~R. Coleman and E.~J. Weinberg, \emph{{Radiative Corrections as the Origin of
  Spontaneous Symmetry Breaking}},
  \href{https://doi.org/10.1103/PhysRevD.7.1888}{\emph{Phys. Rev. D} {\bfseries
  7} (1973) 1888}.

\bibitem{Fendley:1987ef}
P.~Fendley, \emph{{The Effective Potential and the Coupling Constant at High
  Temperature}},
  \href{https://doi.org/10.1016/0370-2693(87)90599-5}{\emph{Phys. Lett. B}
  {\bfseries 196} (1987) 175}.

\bibitem{Parwani:1991gq}
R.~R. Parwani, \emph{{Resummation in a hot scalar field theory}},
  \href{https://doi.org/10.1103/PhysRevD.45.4695}{\emph{Phys. Rev. D}
  {\bfseries 45} (1992) 4695}
  [\href{https://arxiv.org/abs/hep-ph/9204216}{{\ttfamily hep-ph/9204216}}].

\bibitem{Arnold:1992rz}
P.~B. Arnold and O.~Espinosa, \emph{{The Effective potential and first order
  phase transitions: Beyond leading-order}},
  \href{https://doi.org/10.1103/PhysRevD.47.3546}{\emph{Phys. Rev. D}
  {\bfseries 47} (1993) 3546}
  [\href{https://arxiv.org/abs/hep-ph/9212235}{{\ttfamily hep-ph/9212235}}].

\bibitem{Cline:2008hr}
J.~M. Cline, M.~Jarvinen and F.~Sannino, \emph{{The Electroweak Phase
  Transition in Nearly Conformal Technicolor}},
  \href{https://doi.org/10.1103/PhysRevD.78.075027}{\emph{Phys. Rev. D}
  {\bfseries 78} (2008) 075027}
  [\href{https://arxiv.org/abs/0808.1512}{{\ttfamily 0808.1512}}].

\bibitem{Mazumdar:2018dfl}
A.~Mazumdar and G.~White, \emph{{Review of cosmic phase transitions: their
  significance and experimental signatures}},
  \href{https://doi.org/10.1088/1361-6633/ab1f55}{\emph{Rept. Prog. Phys.}
  {\bfseries 82} (2019) 076901}
  [\href{https://arxiv.org/abs/1811.01948}{{\ttfamily 1811.01948}}].

\bibitem{Hindmarsh:2020hop}
M.~B. Hindmarsh, M.~L\"uben, J.~Lumma and M.~Pauly, \emph{{Phase transitions in
  the early universe}},
  \href{https://doi.org/10.21468/SciPostPhysLectNotes.24}{\emph{SciPost Phys.
  Lect. Notes} {\bfseries 24} (2021) 1}
  [\href{https://arxiv.org/abs/2008.09136}{{\ttfamily 2008.09136}}].

\bibitem{Linde:1980tt}
A.~D. Linde, \emph{{Fate of the False Vacuum at Finite Temperature: Theory and
  Applications}},
  \href{https://doi.org/10.1016/0370-2693(81)90281-1}{\emph{Phys. Lett. B}
  {\bfseries 100} (1981) 37}.

\bibitem{Jinno:2016knw}
R.~Jinno and M.~Takimoto, \emph{{Probing a classically conformal B-L model with
  gravitational waves}},
  \href{https://doi.org/10.1103/PhysRevD.95.015020}{\emph{Phys. Rev. D}
  {\bfseries 95} (2017) 015020}
  [\href{https://arxiv.org/abs/1604.05035}{{\ttfamily 1604.05035}}].

\bibitem{Iso:2009nw}
S.~Iso, N.~Okada and Y.~Orikasa, \emph{{The minimal B-L model naturally
  realized at TeV scale}},
  \href{https://doi.org/10.1103/PhysRevD.80.115007}{\emph{Phys. Rev. D}
  {\bfseries 80} (2009) 115007}
  [\href{https://arxiv.org/abs/0909.0128}{{\ttfamily 0909.0128}}].

\bibitem{Ellis:2018mja}
J.~Ellis, M.~Lewicki and J.~M. No, \emph{{On the Maximal Strength of a
  First-Order Electroweak Phase Transition and its Gravitational Wave Signal}},
   \href{https://arxiv.org/abs/1809.08242}{{\ttfamily 1809.08242}}.

\bibitem{Ellis:2020nnr}
J.~Ellis, M.~Lewicki and V.~Vaskonen, \emph{{Updated predictions for
  gravitational waves produced in a strongly supercooled phase transition}},
  \href{https://doi.org/10.1088/1475-7516/2020/11/020}{\emph{JCAP} {\bfseries
  11} (2020) 020} [\href{https://arxiv.org/abs/2007.15586}{{\ttfamily
  2007.15586}}].

\bibitem{Athron:2020sbe}
P.~Athron, C.~Bal\'azs, A.~Fowlie and Y.~Zhang, \emph{{PhaseTracer: tracing
  cosmological phases and calculating transition properties}},
  \href{https://doi.org/10.1140/epjc/s10052-020-8035-2}{\emph{Eur. Phys. J. C}
  {\bfseries 80} (2020) 567}
  [\href{https://arxiv.org/abs/2003.02859}{{\ttfamily 2003.02859}}].

\bibitem{Turner:1990rc}
M.~S. Turner and F.~Wilczek, \emph{{Relic gravitational waves and extended
  inflation}}, \href{https://doi.org/10.1103/PhysRevLett.65.3080}{\emph{Phys.
  Rev. Lett.} {\bfseries 65} (1990) 3080}.

\bibitem{Kosowsky:1991ua}
A.~Kosowsky, M.~S. Turner and R.~Watkins, \emph{{Gravitational radiation from
  colliding vacuum bubbles}},
  \href{https://doi.org/10.1103/PhysRevD.45.4514}{\emph{Phys. Rev. D}
  {\bfseries 45} (1992) 4514}.

\bibitem{Kosowsky:1992rz}
A.~Kosowsky, M.~S. Turner and R.~Watkins, \emph{{Gravitational waves from first
  order cosmological phase transitions}},
  \href{https://doi.org/10.1103/PhysRevLett.69.2026}{\emph{Phys. Rev. Lett.}
  {\bfseries 69} (1992) 2026}.

\bibitem{Kosowsky:1992vn}
A.~Kosowsky and M.~S. Turner, \emph{{Gravitational radiation from colliding
  vacuum bubbles: envelope approximation to many bubble collisions}},
  \href{https://doi.org/10.1103/PhysRevD.47.4372}{\emph{Phys. Rev. D}
  {\bfseries 47} (1993) 4372}
  [\href{https://arxiv.org/abs/astro-ph/9211004}{{\ttfamily
  astro-ph/9211004}}].

\bibitem{Turner:1992tz}
M.~S. Turner, E.~J. Weinberg and L.~M. Widrow, \emph{{Bubble nucleation in
  first order inflation and other cosmological phase transitions}},
  \href{https://doi.org/10.1103/PhysRevD.46.2384}{\emph{Phys. Rev. D}
  {\bfseries 46} (1992) 2384}.

\bibitem{Hindmarsh:2013xza}
M.~Hindmarsh, S.~J. Huber, K.~Rummukainen and D.~J. Weir, \emph{{Gravitational
  waves from the sound of a first order phase transition}},
  \href{https://doi.org/10.1103/PhysRevLett.112.041301}{\emph{Phys. Rev. Lett.}
  {\bfseries 112} (2014) 041301}
  [\href{https://arxiv.org/abs/1304.2433}{{\ttfamily 1304.2433}}].

\bibitem{Giblin:2014qia}
J.~T. Giblin and J.~B. Mertens, \emph{{Gravitional radiation from first-order
  phase transitions in the presence of a fluid}},
  \href{https://doi.org/10.1103/PhysRevD.90.023532}{\emph{Phys. Rev. D}
  {\bfseries 90} (2014) 023532}
  [\href{https://arxiv.org/abs/1405.4005}{{\ttfamily 1405.4005}}].

\bibitem{Hindmarsh:2015qta}
M.~Hindmarsh, S.~J. Huber, K.~Rummukainen and D.~J. Weir, \emph{{Numerical
  simulations of acoustically generated gravitational waves at a first order
  phase transition}},
  \href{https://doi.org/10.1103/PhysRevD.92.123009}{\emph{Phys. Rev. D}
  {\bfseries 92} (2015) 123009}
  [\href{https://arxiv.org/abs/1504.03291}{{\ttfamily 1504.03291}}].

\bibitem{Hindmarsh:2017gnf}
M.~Hindmarsh, S.~J. Huber, K.~Rummukainen and D.~J. Weir, \emph{{Shape of the
  acoustic gravitational wave power spectrum from a first order phase
  transition}}, \href{https://doi.org/10.1103/PhysRevD.96.103520}{\emph{Phys.
  Rev. D} {\bfseries 96} (2017) 103520}
  [\href{https://arxiv.org/abs/1704.05871}{{\ttfamily 1704.05871}}].

\bibitem{Kamionkowski:1993fg}
M.~Kamionkowski, A.~Kosowsky and M.~S. Turner, \emph{{Gravitational radiation
  from first order phase transitions}},
  \href{https://doi.org/10.1103/PhysRevD.49.2837}{\emph{Phys. Rev. D}
  {\bfseries 49} (1994) 2837}
  [\href{https://arxiv.org/abs/astro-ph/9310044}{{\ttfamily
  astro-ph/9310044}}].

\bibitem{Kosowsky:2001xp}
A.~Kosowsky, A.~Mack and T.~Kahniashvili, \emph{{Gravitational radiation from
  cosmological turbulence}},
  \href{https://doi.org/10.1103/PhysRevD.66.024030}{\emph{Phys. Rev. D}
  {\bfseries 66} (2002) 024030}
  [\href{https://arxiv.org/abs/astro-ph/0111483}{{\ttfamily
  astro-ph/0111483}}].

\bibitem{Caprini:2006jb}
C.~Caprini and R.~Durrer, \emph{{Gravitational waves from stochastic
  relativistic sources: Primordial turbulence and magnetic fields}},
  \href{https://doi.org/10.1103/PhysRevD.74.063521}{\emph{Phys. Rev.}
  {\bfseries D74} (2006) 063521}
  [\href{https://arxiv.org/abs/astro-ph/0603476}{{\ttfamily
  astro-ph/0603476}}].

\bibitem{Gogoberidze:2007an}
G.~Gogoberidze, T.~Kahniashvili and A.~Kosowsky, \emph{{The Spectrum of
  Gravitational Radiation from Primordial Turbulence}},
  \href{https://doi.org/10.1103/PhysRevD.76.083002}{\emph{Phys. Rev. D}
  {\bfseries 76} (2007) 083002}
  [\href{https://arxiv.org/abs/0705.1733}{{\ttfamily 0705.1733}}].

\bibitem{Caprini:2009yp}
C.~Caprini, R.~Durrer and G.~Servant, \emph{{The stochastic gravitational wave
  background from turbulence and magnetic fields generated by a first-order
  phase transition}},
  \href{https://doi.org/10.1088/1475-7516/2009/12/024}{\emph{JCAP} {\bfseries
  0912} (2009) 024} [\href{https://arxiv.org/abs/0909.0622}{{\ttfamily
  0909.0622}}].

\bibitem{Niksa:2018ofa}
P.~Niksa, M.~Schlederer and G.~Sigl, \emph{{Gravitational Waves produced by
  Compressible MHD Turbulence from Cosmological Phase Transitions}},
  \href{https://doi.org/10.1088/1361-6382/aac89c}{\emph{Class. Quant. Grav.}
  {\bfseries 35} (2018) 144001}
  [\href{https://arxiv.org/abs/1803.02271}{{\ttfamily 1803.02271}}].

\bibitem{Caprini:2019egz}
C.~Caprini et~al., \emph{{Detecting gravitational waves from cosmological phase
  transitions with LISA: an update}},
  \href{https://doi.org/10.1088/1475-7516/2020/03/024}{\emph{JCAP} {\bfseries
  03} (2020) 024} [\href{https://arxiv.org/abs/1910.13125}{{\ttfamily
  1910.13125}}].

\bibitem{Caprini:2015zlo}
C.~Caprini et~al., \emph{{Science with the space-based interferometer eLISA.
  II: Gravitational waves from cosmological phase transitions}},
  \href{https://doi.org/10.1088/1475-7516/2016/04/001}{\emph{JCAP} {\bfseries
  1604} (2016) 001} [\href{https://arxiv.org/abs/1512.06239}{{\ttfamily
  1512.06239}}].

\bibitem{Cai:2017tmh}
R.-G. Cai, M.~Sasaki and S.-J. Wang, \emph{{The gravitational waves from the
  first-order phase transition with a dimension-six operator}},
  \href{https://doi.org/10.1088/1475-7516/2017/08/004}{\emph{JCAP} {\bfseries
  08} (2017) 004} [\href{https://arxiv.org/abs/1707.03001}{{\ttfamily
  1707.03001}}].

\bibitem{Steinhardt:1981ct}
P.~J. Steinhardt, \emph{{Relativistic Detonation Waves and Bubble Growth in
  False Vacuum Decay}},
  \href{https://doi.org/10.1103/PhysRevD.25.2074}{\emph{Phys. Rev.} {\bfseries
  D25} (1982) 2074}.

\bibitem{Espinosa:2010hh}
J.~R. Espinosa, T.~Konstandin, J.~M. No and G.~Servant, \emph{{Energy Budget of
  Cosmological First-order Phase Transitions}},
  \href{https://doi.org/10.1088/1475-7516/2010/06/028}{\emph{JCAP} {\bfseries
  06} (2010) 028} [\href{https://arxiv.org/abs/1004.4187}{{\ttfamily
  1004.4187}}].

\bibitem{Huber:2013kj}
S.~J. Huber and M.~Sopena, \emph{{An efficient approach to electroweak bubble
  velocities}},  \href{https://arxiv.org/abs/1302.1044}{{\ttfamily 1302.1044}}.

\bibitem{Leitao:2014pda}
L.~Leitao and A.~Megevand, \emph{{Hydrodynamics of phase transition fronts and
  the speed of sound in the plasma}},
  \href{https://doi.org/10.1016/j.nuclphysb.2014.12.008}{\emph{Nucl. Phys. B}
  {\bfseries 891} (2015) 159}
  [\href{https://arxiv.org/abs/1410.3875}{{\ttfamily 1410.3875}}].

\bibitem{Dorsch:2018pat}
G.~C. Dorsch, S.~J. Huber and T.~Konstandin, \emph{{Bubble wall velocities in
  the Standard Model and beyond}},
  \href{https://doi.org/10.1088/1475-7516/2018/12/034}{\emph{JCAP} {\bfseries
  12} (2018) 034} [\href{https://arxiv.org/abs/1809.04907}{{\ttfamily
  1809.04907}}].

\bibitem{Cline:2020jre}
J.~M. Cline and K.~Kainulainen, \emph{{Electroweak baryogenesis at high bubble
  wall velocities}},
  \href{https://doi.org/10.1103/PhysRevD.101.063525}{\emph{Phys. Rev. D}
  {\bfseries 101} (2020) 063525}
  [\href{https://arxiv.org/abs/2001.00568}{{\ttfamily 2001.00568}}].

\bibitem{Lewicki:2021pgr}
M.~Lewicki, M.~Merchand and M.~Zych, \emph{{Electroweak bubble wall expansion:
  gravitational waves and baryogenesis in Standard Model-like thermal plasma}},
  \href{https://doi.org/10.1007/JHEP02(2022)017}{\emph{JHEP} {\bfseries 02}
  (2022) 017} [\href{https://arxiv.org/abs/2111.02393}{{\ttfamily
  2111.02393}}].

\bibitem{McLaughlin:2013ira}
M.~A. McLaughlin, \emph{{The North American Nanohertz Observatory for
  Gravitational Waves}},
  \href{https://doi.org/10.1088/0264-9381/30/22/224008}{\emph{Class. Quant.
  Grav.} {\bfseries 30} (2013) 224008}
  [\href{https://arxiv.org/abs/1310.0758}{{\ttfamily 1310.0758}}].

\bibitem{Weltman:2018zrl}
A.~Weltman et~al., \emph{{Fundamental physics with the Square Kilometre
  Array}}, \href{https://doi.org/10.1017/pasa.2019.42}{\emph{Publ. Astron. Soc.
  Austral.} {\bfseries 37} (2020) e002}
  [\href{https://arxiv.org/abs/1810.02680}{{\ttfamily 1810.02680}}].

\bibitem{Garcia-Bellido:2021zgu}
J.~Garcia-Bellido, H.~Murayama and G.~White, \emph{{Exploring the Early
  Universe with Gaia and THEIA}},
  \href{https://arxiv.org/abs/2104.04778}{{\ttfamily 2104.04778}}.

\bibitem{Sesana:2019vho}
A.~Sesana et~al., \emph{{Unveiling the gravitational universe at $\mu$-Hz
  frequencies}}, \href{https://doi.org/10.1007/s10686-021-09709-9}{\emph{Exper.
  Astron.} {\bfseries 51} (2021) 1333}
  [\href{https://arxiv.org/abs/1908.11391}{{\ttfamily 1908.11391}}].

\bibitem{AmaroSeoane2012LaserIS}
P.~Amaro-Seoane, H.~Audley, S.~Babak, J.~M. Baker, E.~Barausse, P.~L. Bender
  et~al., \emph{Laser interferometer space antenna},  2012.

\bibitem{Kawamura:2006up}
S.~Kawamura et~al., \emph{{The Japanese space gravitational wave antenna
  DECIGO}}, \href{https://doi.org/10.1088/0264-9381/23/8/S17}{\emph{Class.
  Quant. Grav.} {\bfseries 23} (2006) S125}.

\bibitem{Yagi:2011wg}
K.~Yagi and N.~Seto, \emph{{Detector configuration of DECIGO/BBO and
  identification of cosmological neutron-star binaries}},
  \href{https://doi.org/10.1103/PhysRevD.83.044011}{\emph{Phys. Rev. D}
  {\bfseries 83} (2011) 044011}
  [\href{https://arxiv.org/abs/1101.3940}{{\ttfamily 1101.3940}}].

\bibitem{Punturo_2010}
M.~Punturo, M.~Abernathy, F.~Acernese, B.~Allen, N.~Andersson, K.~Arun et~al.,
  \emph{The einstein telescope: a third-generation gravitational wave
  observatory},
  \href{https://doi.org/10.1088/0264-9381/27/19/194002}{\emph{Classical and
  Quantum Gravity} {\bfseries 27} (2010) 194002}.

\bibitem{LIGOScientific:2016wof}
{\scshape LIGO Scientific} collaboration, \emph{{Exploring the Sensitivity of
  Next Generation Gravitational Wave Detectors}},
  \href{https://doi.org/10.1088/1361-6382/aa51f4}{\emph{Class. Quant. Grav.}
  {\bfseries 34} (2017) 044001}
  [\href{https://arxiv.org/abs/1607.08697}{{\ttfamily 1607.08697}}].

\bibitem{LIGOScientific:2014pky}
{\scshape LIGO Scientific} collaboration, \emph{{Advanced LIGO}},
  \href{https://doi.org/10.1088/0264-9381/32/7/074001}{\emph{Class. Quant.
  Grav.} {\bfseries 32} (2015) 074001}
  [\href{https://arxiv.org/abs/1411.4547}{{\ttfamily 1411.4547}}].

\bibitem{Planck:2018vyg}
{\scshape Planck} collaboration, \emph{{Planck 2018 results. VI. Cosmological
  parameters}},
  \href{https://doi.org/10.1051/0004-6361/201833910}{\emph{Astron. Astrophys.}
  {\bfseries 641} (2020) A6}
  [\href{https://arxiv.org/abs/1807.06209}{{\ttfamily 1807.06209}}].

\bibitem{Giudice:2003jh}
G.~Giudice, A.~Notari, M.~Raidal, A.~Riotto and A.~Strumia, \emph{{Towards a
  complete theory of thermal leptogenesis in the SM and MSSM}},
  \href{https://doi.org/10.1016/j.nuclphysb.2004.02.019}{\emph{Nucl. Phys. B}
  {\bfseries 685} (2004) 89}
  [\href{https://arxiv.org/abs/hep-ph/0310123}{{\ttfamily hep-ph/0310123}}].

\bibitem{Casas:2001sr}
J.~A. Casas and A.~Ibarra, \emph{{Oscillating neutrinos and $\mu \to e,
  \gamma$}}, \href{https://doi.org/10.1016/S0550-3213(01)00475-8}{\emph{Nucl.
  Phys. B} {\bfseries 618} (2001) 171}
  [\href{https://arxiv.org/abs/hep-ph/0103065}{{\ttfamily hep-ph/0103065}}].

\bibitem{Gonzalez-Garcia:2021dve}
M.~C. Gonzalez-Garcia, M.~Maltoni and T.~Schwetz, \emph{{NuFIT: Three-Flavour
  Global Analyses of Neutrino Oscillation Experiments}},
  \href{https://doi.org/10.3390/universe7120459}{\emph{Universe} {\bfseries 7}
  (2021) 459} [\href{https://arxiv.org/abs/2111.03086}{{\ttfamily
  2111.03086}}].

\bibitem{Hall:2009bx}
L.~J. Hall, K.~Jedamzik, J.~March-Russell and S.~M. West, \emph{{Freeze-In
  Production of FIMP Dark Matter}},
  \href{https://doi.org/10.1007/JHEP03(2010)080}{\emph{JHEP} {\bfseries 03}
  (2010) 080} [\href{https://arxiv.org/abs/0911.1120}{{\ttfamily 0911.1120}}].

\bibitem{Belanger:2014vza}
G.~Bélanger, F.~Boudjema, A.~Pukhov and A.~Semenov, \emph{{micrOMEGAs4.1: two
  dark matter candidates}},
  \href{https://doi.org/10.1016/j.cpc.2015.03.003}{\emph{Comput. Phys. Commun.}
  {\bfseries 192} (2015) 322}
  [\href{https://arxiv.org/abs/1407.6129}{{\ttfamily 1407.6129}}].

\bibitem{Bhattacharya:2019tqq}
S.~Bhattacharya, N.~Chakrabarty, R.~Roshan and A.~Sil, \emph{{Multicomponent
  dark matter in extended $U(1)_{B-L}$: neutrino mass and high scale
  validity}},  \href{https://arxiv.org/abs/1910.00612}{{\ttfamily 1910.00612}}.

\bibitem{Adams:1993zs}
F.~C. Adams, \emph{{General solutions for tunneling of scalar fields with
  quartic potentials}},
  \href{https://doi.org/10.1103/PhysRevD.48.2800}{\emph{Phys. Rev. D}
  {\bfseries 48} (1993) 2800}
  [\href{https://arxiv.org/abs/hep-ph/9302321}{{\ttfamily hep-ph/9302321}}].

\end{thebibliography}

\providecommand{\href}[2]{#2}\begingroup\raggedright\endgroup

\end{document}